\pdfoutput=1
\documentclass[conference,twocolumn]{IEEEtran}

\usepackage{multicol}
\usepackage[nolist,nohyperlinks]{acronym}
\usepackage{mathtools}
\usepackage{breqn}
\usepackage{amsmath}
\usepackage{balance}
\usepackage{framed}
\usepackage{siunitx}
\usepackage{soul}
\usepackage{multirow}
\usepackage{etoolbox}
\usepackage{verbatim}
\usepackage{graphicx}
\usepackage{subcaption}
\usepackage{color}
\usepackage[table,xcdraw]{xcolor}
\usepackage{float}
\graphicspath{{./images/}}

\usepackage{url}
\newtoggle{authorcomment}
\toggletrue{authorcomment}

\usepackage{algorithm}
\usepackage{algorithmic}

\begin{document}
% Control the number of names in references
\bstctlcite{IEEEexample:BSTcontrol}

\title{Adaptive Height Optimisation for Cellular-Connected UAVs: A Deep Reinforcement Learning Approach}

% Author block
\author{
  \IEEEauthorblockN{
    Erika Fonseca*,
    Boris Galkin*,
    Ramy Amer*,
    Luiz A. DaSilva$^\ddag$,
    Ivana Dusparic*
  }\\
  \IEEEauthorblockA{
    \IEEEauthorrefmark{1} CONNECT - Trinity College Dublin, Ireland,\\
    $^\ddag$Commonwealth Cyber Initiative, Virginia Tech, USA \\
    \{fonsecae, galkinb, ramyr, duspari\}@tcd.ie,  ldasilva@vt.edu}\\
    
% \IEEEauthorblockA{
%    \IEEEauthorrefmark{2} Some Nice University, Country,
%    \{thing\}@somewhere.com}\\
% \textit{“This work has been submitted to the IEEE for possible publication. Copyright may be transferred without notice, after which this version may no longer be accessible.”}
  }

\maketitle
% Load acronym file
% !TEX root = ../main.tex
%\section*{Acronyms}

\begin{acronym}[IMT-Advanced]
  \acro{3gpp}[3GPP]{3\textsuperscript{rd} Generation Partnership Program}
\acro{5g}[5G]{Fifth Generation Mobile Networks}
\acro{Adam}{Adaptive Moment Optimisation}
%\acro{AdaMax}{Adaptive Moment Optimisation}
\acro{ap}[AP]{Access Point}
\acro{atc}[ATC]{Air Traffic Control}
\acro{bs}[BS]{Base Station}
\acro{bpp}[BPP]{Binomial Point Process}
\acro{cc}[C\&C]{Command \& Control}
\acro{cdf}[CDF]{cumulative distribution function}
\acro{cdma}[CDMA]{Code Division Multiple Access}
\acro{cdr}[CD\&R]{Conflict Detection \& Resolution}
\acro{cfo}[CFO]{Carrier Frequency Offset}
\acro{comreg}[ComReg]{Commission for Communications Regulation}
\acro{csi}[CSI]{Channel State Information}
\acro{easa}[EASA]{European Aviation Safety Agency}
\acro{dql}[DQN]{Deep Q-Learning}
\acro{esn}[ESN]{echo state network}
\acro{faa}[FAA]{Federal Aviation Administration}
\acro{fso}[FSO]{Free-Space Optical}
\acro{geo}[GEO]{Geosynchronous Equatorial Orbit}
\acro{gps}[GPS]{Global Positioning System}
\acro{gs}[GS]{Ground Station}
\acro{gue}[GUE]{Ground User Equipment}
\acro{hap}[HAP]{High Altitude Platform}
\acro{iot}[IoT]{Internet of Things}
\acro{kpi}[KPI]{Key Performance Indicator}
\acro{lap}[LAP]{Low Altitude Platform}
\acro{leo}[LEO]{Low Earth Orbit}
\acro{los}[LoS]{Line-of-Sight}
\acro{lte}[LTE]{Long Term Evolution}
\acro{mac}[MAC]{Media Access Control}
\acro{mcp}[MCP]{Matern Cluster Process}
\acro{mc}[MC]{Monte Carlo}
\acro{mimo}[MIMO]{Multiple Input Multiple Output}
\acro{mip}[MIP]{Mixed-Integer Programming}
\acro{milp}[MILP]{Mixed-Integer Linear Programming}
\acro{mm}[MM]{Mapping Mechanism}
\acro{ml}[ML]{machine learning}
\acro{mno}[MNO]{Mobile Network Operator}
\acro{nlos}[NLoS]{non-Line-of-Sight}
\acro{nn}[NN]{Neural Network}
\acro{ofdma}[OFDMA]{Orthogonal Frequency Division Multiple Access}
\acro{oot}[OOT]{Out-of-Tree}
\acro{osi}[OSI]{Open Systems Interconnection}
\acro{otdoa}[OTDoA]{Observed Time Difference of Arrival}
\acro{ott}[OTT]{Over-The-Top}
\acro{pv}[PV]{photo-voltaic}
\acro{pdf}[pdf]{probability density function}
\acro{ppp}[PPP]{Poisson Point Process}
\acro{qos}[QoS]{Quality of Service}
\acro{rc}[RC]{Remote Control}
\acro{rl}[RL]{Reinforcement Learning}
 \acro{dql}[DQN]{Deep Q-Learning}
 \acro{DL}[DNN]{Deep Neural Network}
\acro{rss}[RSS]{Received Signal Strength}
\acro{se}[SE]{Spectral Efficiency}
\acro{sir}[SIR]{Signal-to-Interference Ratio}
\acro{SINR}[SINR]{Signal-to-Interference-and-Noise Ratio}
\acro{snr}[SNR]{Signal-to-Noise Ratio}
\acro{uav}[UAV]{Unmanned Aerial Vehicle}
\acro{ue}[UE]{User Equipment}
\acro{ula}[ULA]{Uniform Linear Array}
\acro{wsn}[WSN]{Wireless Sensor Network}
\acro{wsn}[WSN]{Wireless Sensor Network}
\acro{rv}[RV]{random variable}
\acro{ppp}[PPP]{Poisson point process}
\acro{pgfl}[PGFL]{point generation functional}
\acro{pdf}[PDF]{probability density function}
\acro{pso}[PSO]{particle swarm optimisation}
\acro{5G}{fifth generation of wireless technology}
\acro{3GPP}{3rd Generation Partnership Project }

\end{acronym}
% Abstract
% that in the physical layer is related to its spectrum efficiency.
\begin{abstract}
Providing reliable connectivity to cellular-connected \acp{uav} can be very challenging; their performance highly depends on the nature of the surrounding environment, such as density and heights of the ground \acp{bs}. On the other hand, tall buildings might block undesired interference signals from ground \acp{bs}, thereby improving the connectivity between the \acp{uav} and their serving \acp{bs}. To address the connectivity of \acp{uav} in such environments, this paper proposes a \ac{rl} algorithm to dynamically optimise the height of a \ac{uav} as it moves through the environment, with the goal of increasing the throughput or spectrum efficiency that it experiences. The proposed solution is evaluated in two settings: using a series of generated environments where we vary the number of \ac{bs} and building densities, and in a scenario using real-world data obtained from an experiment in Dublin, Ireland. 
Results show that our proposed RL-based solution improves \ac{uav} \ac{qos} by 6\% to 41\%, depending on the scenario. We also conclude that, when flying at heights higher than the buildings, building density variation has no impact on UAV QoS. On the other hand, BS density can negatively impact UAV QoS, with higher numbers of BSs generating more interference and deteriorating UAV performance.

%is capable of exceeding the benchmarks from 41\% to 5\% of its throughput. 

\end{abstract}

\begin{IEEEkeywords}
Unmanned Aerial Vehicles (UAVs), Reinforcement Learning, Two-tier networks, Experimental Measurements, Massive MIMO.
\end{IEEEkeywords}

\IEEEpeerreviewmaketitle

% Reset acronyms after title and abstract
\acresetall

% Load documents sections
\section{Introduction}
\label{sec:intro}
% !TEX root = ../../main.tex

%Relevancy of drones in the society, What is the challenge? Who cares?
\acp{uav} can leverage 5G connectivity to perform different applications, such as security surveillance, search and rescue operations, and building inspections. However, providing reliable connectivity to such \acp{uav} is still an open problem, as they present a paradigm shift when compared to their ground counterparts such as smartphones. 
According to the specifications (release 14 of \ac{3GPP} \cite{release14}), a \ac{uav} needs to maintain continuous connectivity with the mobile network at speeds up to 300km/h.

%Is it possible to use actual infra?
Previous work, such as \cite{rami1, rami2}, investigates the feasibility of using existing network infrastructure to provide reliable wireless connectivity for \acp{uav}. These studies conclude that currently deployed networks would need to adapt some of their design configurations, such as increasing \ac{bs} heights \cite{reshape} or changing the tilt of the antennas \cite{network-adapt} so as to enable connectivity for \acp{uav}. Redesigning the terrestrial network infrastructure may be unfeasible, and an adaptable solution on the \ac{uav} side may be necessary to accelerate the \ac{uav} integration into the network. 

%A bit of what people did to introduce our specific problem
Due to the height at which \acp{uav} fly, there are often no obstacles and therefore no blockage between the \acp{uav} and their serving \ac{bs}. 
However, at high altitudes, the increased probability of \ac{los} to ground \acp{bs} results in high levels of interference at the \acp{uav}.
The work in \cite{reshape} states that the optimal height at which the \ac{uav} can fly to maintain reliable communication depends on the \ac{bs} density and height. Similarly, the authors in \cite{rami3} show that the vertical movements of the \ac{uav} affect their coverage probability.
%They evaluated how the BS height and antenna angle influences the coverage for \ac{gue} and \acp{uav} at 100m altitude. It shows that for a BS density of 100 BS/Km$^2$ there is no coverage to the \ac{uav}; however, it can have coverage in case of 1 and 10 BS/Km$^2$. 
%A \ac{uav} can fly over different environments where the local BS density can change, which will force it to change its height to optimise its wireless link to the network, with respect to the local environmental conditions.

%To date, no other work addresses the dynamic height optimisation of \acp{uav} acting as end-users of the cellular network. 
%Adaptive height optimisation has only been discussed in the scenario where \acp{uav} act as \acp{bs}, supplementing the coverage of \ac{gue} in the mobile network. %The problem is considerably different as they are stationary while in our scenario, the \ac{uav} moves through different environments. 
Motivated by the above, this paper proposes a \ac{rl} approach for dynamic optimisation of the height of a \ac{uav} connected to the cellular network once it moves through a city.
We propose optimising the altitude of the \ac{uav}, separating it from the problem of a horizontal trajectory decision. We separate it from the horizontal optimisation trajectory as in some applications, such as surveillance and organ delivery, the horizontal path will be defined by the application, and only the altitude will have the freedom to be adapted.
We evaluate our proposed approach with generated environment and a experimental measurement dataset.
We investigate the proposed solution in a generated environment to evaluate which are the main characteristics to influence the approach. In this environment, we vary the \ac{bs} and building densities to understand if these variables interfere with the optimal \ac{uav} altitude.

Then to complement the investigation we adapt the proposed approach to be used with data collected from real-world scenario. 
%The real-world dataset was collected with a smartphone attached to a \ac{uav} that flew in two different areas of Dublin city centre. The \ac{uav} is connected to a macro cell and a small cell network in the first and second scenarios, respectively.
To  the best  of  our  knowledge,  this  is  the  first  work  to optimise connectivity of a cellular-connected UAV by dynamically adapting the height at which it is flying, as well as the first to evaluate a UAV connectivity optimisation approach on experimentally-obtained real-world data. 
The main contributions of this paper are described bellow:
\begin{itemize}
  \item We define a problem of optimising \ac{qos} of an \ac{uav} as an \ac{rl} problem, defining states and actions.
  \item We propose a solution to adapt the \ac{uav}'s height dynamically, which uses \ac{dql} and replay memory to optimise \ac{qos} parameters as the spectrum efficiency and throughput.
  \item We provide an evaluation of the influence of building density on the \ac{uav} height adaptation.
  \item We provide an evaluation of the proposed solution in a generated environment and using a real-world based dataset.
  \item We analyse how the proposed solution and the baselines affect the height. 
  %\item The proposed height adaptation solution can work with different approaches that define the horizontal path, being easy to adapt to applications such as organ delivery and surveillance.
  %\item We are the first to analyse \ac{uav} dynamically height optimisation.
  %\item We are the first to analyse the impact of building density in the \ac{uav}'s height.
  %\item We are the first to use experimental data in the evaluation of connected \ac{uav}s.
\end{itemize}

\iffalse
Our proposed \ac{uav} height optimisation could be used with the following purposes:

\begin{itemize}
    \item Be used by the \ac{uav} to take informed decisions of its next height in order to improve its \ac{qos}.
    \item Be used to automatically avoid areas with poor signal in the sky.
    \item Be used by different \ac{uav} applications as it does not require modification of the horizontal path. 
\end{itemize}
\fi
%Explaining the rest of the paper
The remainder of the paper is organised as follows. In Section \ref{sec:related} we discuss the existing work done on the issue of connectivity of \acp{uav} to the wireless network. In Section \ref{sec:systemmodel} we present the system model of our generated environment. In Section \ref{sec:problem} we introduce the 
problem statement, where we define the scenario and how the \ac{uav} moves. In Section \ref{sec:ml} the design and implementation of our proposed \ac{rl} solution is explained. 
We detail the parameters of our RL model, as well as the algorithm.  %Section \ref{sec:experiment} details the experimental setup of a connected \ac{uav} flying in two different locations of Dublin city centre and the simulation description with varying BS and building densities.
In Section \ref{sec:evaluation}, we  evaluate our solution for the scenario where we use generated data.  
In Section \ref{sec:data-exp} we introduce the real-world dataset and detail small changes on the proposed solution to use this data, then in Section \ref{sec:evaluation-exp}, we present the results using the real-world dataset.
Finally, in Section \ref{sec:conclusion}, we conclude the paper and discuss the issues that remain open for future work.

\section{UAV movement optimisation: Related Work}
\label{sec:related}
The works on trajectory optimisation focus on 2D optimisation and rarely mention the height or the \ac{uav}. In this section, we introduce works that optimise the trajectory considering the \ac{uav}-\ac{bs} access link.

\begin{table*}[h]
\centering
\begin{tabular}{|c|c|c|c|}
\hline
\rowcolor[HTML]{C0C0C0} 
\multicolumn{1}{|c|}{\cellcolor[HTML]{C0C0C0}\textbf{Paper}} &
\multicolumn{1}{|c|}{\cellcolor[HTML]{C0C0C0}\textbf{Type of UAV}} &
\multicolumn{1}{c|}{\cellcolor[HTML]{C0C0C0}\textbf{Optimise}} &
\multicolumn{1}{c|}{\cellcolor[HTML]{C0C0C0}\textbf{Method}}\\ \hline
\cite{trajectory1} & Connected UAV & Distance to BS & Graph  \\ \hline
\cite{rami3} & connected UAV & Coverage Prediction & Cauchy’s inequality  \\ \hline
\cite{trajec-mobile} & Connected UAV & Horizontal Optimisation & Graph  \\ \hline
\cite{rival} & Connected UAV & Horizontal Optimisation & Deep \ac{rl}  \\ \hline
\cite{3d-2} & UAV BS & 3D position & Bisection search   \\ \hline
\cite{3d-1} & UAV BS & 3D position & Particle swarm optimisation    \\ \hline
\cite{3d-3} & UAV BS & 3D position & \ac{dql}   \\ \hline

%\cite{} & & &  &  &   \\ \hline

\end{tabular}
\caption{
UAV movement optimisation works.
	}
	\label{table:uav-works}
\end{table*}

In \cite{trajec-mobile}, the authors propose optimising the horizontal path of a cellular connected \ac{uav} that flies from an initial to a final location, while maintaining reliable communication with the underlying mobile network. This approach proposes that the \ac{uav} flies at the fixed minimum height allowed by the regulatory entities. In this study, the authors do not consider the interference from \acp{bs} to which the \ac{uav} is not connected and blockage from the buildings blocking the link from \ac{uav} to \ac{bs}. To accomplish the study objectives, a graph representation of the network is proposed, with 3 solutions: first, a graph where each node is a \ac{bs}; second, a graph where the nodes are the handover points between the \ac{bs}s; and third, where the handover points are the optimal point in an intersection area. Dijkstra algorithm is used to find the route of the \ac{uav} and show it is close to the optimal solution. 
Height optimisation was not considered, and the authors conclude that introducing a height variable to the problem is not a trivial task and that their proposed horizontal trajectory solution is not the most appropriate one for 3D movement. They conclude that it would be unfeasible to represent all the possible heights a \ac{uav} could have at all the nodes, as each of them should be a new node increasing the system's complexity. 

The work in \cite{rival} creates an optimised path with the objective of maintaining a uninterrupted connection to the \acp{bs}. This work only considers the uplink from the \ac{uav} to the \ac{bs} network. This work also highlights the importance of the altitude of the \ac{uav} and calculates the upper and lower bounds for the height at which the \ac{uav} should fly to satisfy the minimum rate requirements of the uplink, considering the known \ac{bs}s locations. 
Authors calculate a range of heights at which the \ac{uav} should fly to provide a minimum achievable rate. In addition, building blockage on the link \ac{uav} - \ac{bs} is not considered. %The solution is based on the game theory multi-agent approach, where each \ac{uav} is the player. They propose a \ac{ESN} \cite{jaeger2007echo}, which is a recurrent \ac{nn} that where the connectivity and weights of the hidden layers are fixed and randomly assigned.
With this approach, each \ac{uav} decides its next horizontal location. The authors conclude that the altitude is vital to minimise the transmission delay of the \ac{uav} and that it should be a function of the ground network density, network parameters as the transmission power, ground network data requirements and the \ac{uav}'s action. The exact height of the \ac{uav} is not calculated as it would increase the complexity of the algorithm exponentially.

The height planning of a connected \ac{uav} is a new field, however several works have studied the height placement of \ac{uav}s acting as \acp{bs}. The techniques used to optimise the heights at which a \ac{uav} acting as a \ac{bs} should fly can also overlap with our problem of interest as it also consider the radio link between a \ac{uav} and a element that is located at lower heights.

In many examples of the prior art, works on \ac{uav} wireless connectivity, either for \ac{uav} as network end-user or \ac{uav} as \ac{bs}, did not consider the effect of interference conditions. The quality of the link between \ac{uav} and a \ac{bs} can suffer from interference coming from other \ac{bs}s, from objects or buildings intercepting the directional connection between them (shadow zone), or even the natural fading on the propagation. 
The work in \cite{44} assumes \ac{uav} as \ac{bs} and provides coverage to \acp{gue}. The authors propose a sigmoid model to investigate the probability of \ac{los} channel in the \ac{uav} - \ac{gue} link as a function of the vertical angle between them. In the paper, a \ac{uav} with an omnidirectional antenna flies over an urban area. The authors do not consider any source of interference, leaving the link limited with only the path loss.  They conclude that a bigger angle decreases the probability of a building block the link. They also add that there exists an optimal height for the \ac{uav} \ac{bs}, which increases the coverage area. %In the proposed work, we consider not only the building interference, but also the interference generated by other relevant \ac{bs}s to the link \ac{uav}-\ac{gue}.

In \cite{61} and \cite{62}, authors applied stochastic geometry to model the coverage probability of a \ac{uav}-\ac{bs} network in a fading-free and Nakagami-m fading channel. The authors fix the number of \ac{uav}s operating in an area at a certain height above the ground and demonstrate that with an increase in height, the coverage probability decreases.
Also, in \cite{62} authors demonstrate that bigger values of fading parameter reduce the variance of the \ac{SINR} for the \ac{gue}.

In \cite{3d-2} authors propose the approach to calculate the 3D position of \ac{uav} as a \ac{bs}, by applying the interior point optimiser of %MOSEK solver and
bisection search.  Their main objective is to maximise the coverage area by a \ac{uav} cell without providing to a \acs{gue} a \ac{qos} below a threshold. They consider building blockage and \ac{nlos} between the \ac{uav} as \ac{bs} and its users. %, although they do not vary the parameters related to the building density of the urban area in their investigations. %They also vary the quantity of \acp{ue} on their simulations.  
The covered area changes depending on \ac{uav}'s height, and for the lower density of \acp{gue} the coverage is larger when compared to higher density, showing the worst coverage for urban scenarios.

In \cite{3d-1}, the authors find the optimal positions for a network of \ac{uav} as \ac{bs} in order to minimise the number of \acp{bs} required to provide the needed \ac{qos} for their users. The study considers the blockage generated by buildings in an urban area and \ac{nlos} occurrences between the \ac{uav} and its users. %However, authors do not vary the parameters related to the density of the urban area on their investigations. 
The proposed solution used an heuristic algorithm based on the number of \ac{bs} that can serve the \ac{gue}, coverage and capacity requirements. 
The number of users on the network was essential to define the height and number of \acp{uav} as \acp{bs}. The authors concluded that with their solution it is possible to decrease the amount of \ac{uav} as \ac{bs} and provide the same quality on data rate. 
%The authors conclude that by changing the height of the \ac{uav} \acp{bs} they could increase or decrease the overall coverage area.    

Similarly, the work in \cite{3d-3} proposes a 3-step solution for horizontal and vertical optimisation for \ac{uav}-\acp{bs} with different machine learning algorithms for each step. 
The bounds of the \ac{uav} height are the \ac{uav} maximum transmission power for its greatest height, and the minimum required distance between the \ac{uav} and the users, defined by the regulatory entities, for the minimum height. 
In the first instance, it considers a static problem, where the users of the network do not move. As a first step, it partitions the area into cells for the \ac{uav}-\ac{bs}s to cover, by applying K-means (GAK-means) algorithm. Next, it uses a Q-learning algorithm, where each \ac{uav} is an agent and has to decide its position by learning from its mistakes. As the final step, they consider a scenario where users move between \ac{bs}s and the network have to adapt to these movements. 
The authors apply \ac{DL}, as it enables each \ac{uav} to gradually learn the dynamic movements of the users. They conclude that the proposed solution outperforms the K-means algorithm and IGK algorithm with low complexity. 

The  \ac{uav}-\ac{bs} scenario defers from the connected \ac{uav} problem because the connect \ac{uav} moves through the city and do not divide it into cells, so the use of K-mean for clustering, for example, is not applicable. 
However, the use of \ac{rl} to adapt its height depending on the cellular network radio technology and the regulatory entities definitions is a valuable insight. To apply \ac{DL} into the connected \ac{uav} scenario, one needs to investigate what is relevant to a \ac{uav} as \ac{ue}, which are the information a \ac{ue} has from its connection, how the \ac{uav} can interact with the environment, and design a model that can learn all these characteristics and be effective through different topologies.

Table \ref{table:uav-works} provides a summary of the state of the art in connected \ac{uav} movement optimisation.

While some existing work has looked into optimising the height of \ac{uav} \acp{bs}, there is a significant lack of work looking at \ac{uav}s when they are the end users.
In this paper, we propose dynamically optimising the altitude of the \ac{uav}.
The proposed solution applies \ac{rl} to decide, based on environmental measurements, if the \ac{uav} needs to move above, below, or stay at the same height in order to experience the best \ac{qos} possible from the cellular network in the long run.
\section{System Model}
\label{sec:systemmodel}

We consider an urban scenario where a \ac{uav} flies while connected to the cellular network.
The \ac{uav}'s initial and final positions are denoted as $(x_1,y_1,z_1)$ and $(x_f,y_f,z_f)$, with $f$ representing the total number of discrete steps in the experiment. $x$ and $y$ denote coordinates on the horizontal plane, while $z$ denotes height above ground. At each step, the UAV moves in the $x$ coordinate in direction to its final destination.

\subsubsection{Building and BS distribution}
The buildings distributed in the area might affect the \ac{uav} \ac{los}, as they can block the channel between the \ac{uav} and the \ac{bs}s. In order to check if a signal is in \ac{los} or not, we verify if there is a  tall enough building between the \ac{uav} and \ac{bs}. If the signal is blocked by a building, \ac{nlos}, it causes the signal to be attenuated, which is reflected in the \ac{SINR} expression in Equation \ref{eq:SINR}. %We needed to defined this expressions as \ac{los} and \ac{nlos} affect the calculation of the \ac{SINR}. But in real-world implementations there is no need for this information, as the \ac{SINR} is measured.
We use a commonly-adopted model for the urban environment which models the buildings as a square grid with the locations of building centerpoints $(x_{bl},y_{bl})$, that was presented in \cite{ITUR_2012} and used in works as \cite{galkincoverage,3d-1,3d-2}.
The area occupied by each building, $Bl_a$, is constant, and the density of buildings, $Build_{dens}$, is denominated by the number of building per square kilometre. 
The individual building height, $h_bl$  is randomly distributed according to a Poisson distribution, with scale parameter $a$. 

To define the position of the building centerpoints and BSs we run a Poisson distribution with the building and BS densities as input.

\subsubsection{UAV and BS Antennas}
The \ac{uav} is equipped with one omnidirectional antenna to connect to a serving \ac{bs} and receive data. 
The antenna has an omnidirectional radiation pattern, and it has an antenna gain equal to 1. 
We express the coordinates of the \ac{bs} which the \ac{uav} is associated as  $b_s =\{x_s,y_s\} \in \Phi$ and its horizontal distance to the \ac{uav} as $r_s$. The BSs that the UAV is not connected will be called as neighbours BSs.

The \ac{bs} has a directional antenna with a horizontal and vertical beam-width $\omega$ along with a rectangular radiation pattern; The antenna gain is defined as $ \eta(\omega)=0$ outside of the main lobe; and $16\pi/(\omega^2)$ inside of the main lobe. 

Spectrum efficiency $SE$ is the maximum bit rate that can be transmitted per unit of bandwidth. It is a measure of the \ac{qos} in the network. 
%\begin{equation}
% SE = B x \log_2(1+\frac{S}{N})
%\label{eq:SE}
%\end{equation}
The Shannon–Hartley theorem bounds the maximum achievable rate a user can reach once it establishes a wireless link. As we want to improve user's experience providing reliable connectivity to \acp{uav}, our purpose is to increase spectrum efficiency.
We calculate the spectrum efficiency value for the calculated \ac{SINR} based on Shannon–Hartley theorem. 
The \ac{SINR} is a function of the antenna gain and channel model and given as:

\begin{equation}
 SINR = \frac{p \eta(\omega) c (\Delta x^2+\Delta y^2)^{-\alpha_{t_s}/2}}{I_{L} + I_{N}+\sigma^2}
\label{eq:SINR}
\end{equation}

\noindent
 where $p$ is the BS transmit power,  $\alpha_{t_{s}}$ is the pathloss exponent, $t_s \in \{\text{L},\text{N}\}$ indicates whether the \ac{uav} has \ac{los} or \ac{nlos} to its serving BS, $\Delta x^2$ and $\Delta y^2$ is the distance between the BS and the UAV, $c$ is the near-field pathloss, $\sigma^2$ is the noise power, and $I_{L}$ and $I_{N}$ are the aggregate interference from \ac{los} and \ac{nlos}, respectively.

\subsubsection{Horizontal route adaptation}
The \ac{uav} horizontal route is defined by an independent approach that focuses on bringing the \ac{uav} closer to the \ac{bs} it is connected. We introduce this adaptation to the horizontal path so we can investigate the independence of the proposed height adaptation method to the horizontal route. %, as the one described in \cite{trajec-mobile} (we are not applying the same solution on this investigation, but it also have the idea of being closer to the \ac{bs}).
The \ac{uav} flies in direction to its final destination but approximates its Y trajectory to get closer to the \ac{bs} that it is connected by $d$. At every time step, the \ac{uav} connects to the \ac{bs} with stronger \ac{SINR} and get closer in the Y coordinates to this \ac{bs} by $d$, being maximum of $d$ distant to the straight line between $(x_{1},y_{1})$ and $(x_{f},y_{f})$ as illustrated in Figure \ref{fig:nonstraight}.  The focus of our approach is
to investigate if the approach is able to adapt the height of the \ac{uav} and can adapt to any underlying routes decision that a \ac{uav} might take during its path, showing its independence from the horizontal path decisions.

\subsubsection{UAV-BS Link}

The \ac{uav} connects to the \ac{bs} with the best \ac{SINR} at all times.  Therefore, as the \ac{uav} moves through the environment some \acp{bs} become stronger and others weaker. When it reaches the point where its serving \ac{bs} is no longer the \ac{bs} with strongest signal, it will reconnect to the new \ac{los} with the strongest signal. We assume that this handover occurs seamlessly, and there is no disconnect or loss of signal quality when it happens.

We assume that the \ac{uav} will have access to the \ac{SINR} measurements from the \ac{bs} it is connected to and from the 5 neighbours \ac{bs} with strongest signals, the spectrum efficiency it is achieving with the serving \ac{bs}s, and its height at all steps.  \ac{SINR} and spectrum efficiency data is easily obtained by the \ac{uav} from its cellular connection, while the height information is obtained via other \ac{uav} sensors located on the \ac{uav}. 

\section{Problem Statement}
\label{sec:problem}
% !TEX root = ../../main.tex

In this work, as the focus is on \ac{uav} height optimisation and many approaches for optimising 2D trajectories already exist, we assume a simple horizontal path%(ie. either a straight line towards the destination for the experimental analyses, or allowing minor deviations to move closer to serving BS in the simulation analyses)
. Note that simplification of the path does not affect the applicability of our proposed approach, as due to its design, it can be integrated with more complex horizontal path algorithms (which are out of the scope of this paper). In other words, the only coordinate that can be optimised is $z$. We assume that the maximum height change at each time step is $d$, so $|z_t - z_{t-1}| \le d$, where $|.|$ denotes absolute value.

\begin{figure}[t]
  \centering
   \hspace*{\fill}%
  \begin{subfigure}[b]{\columnwidth}
  \centering
    \includegraphics[width=0.7\linewidth]{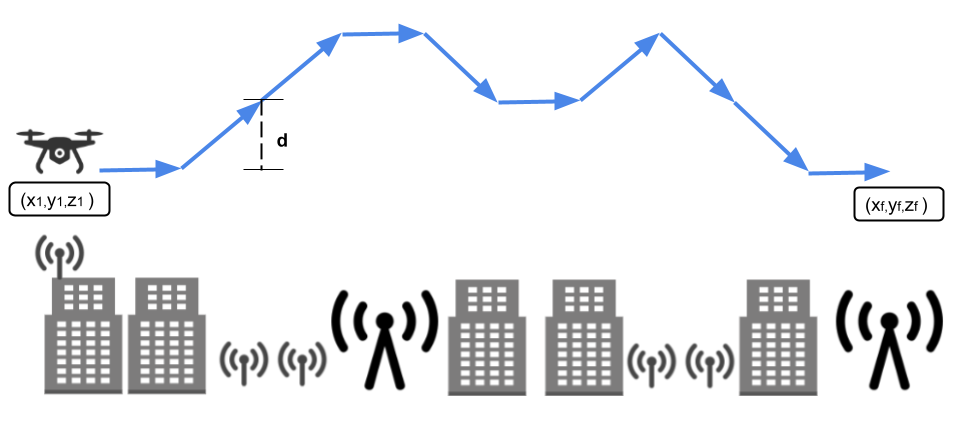}
    \caption{Side view of a UAV connected to the mobile network adjusting  its  height  to maximise its spectrum efficiency.}
    \label{fig:flying-sim}
  \end{subfigure}
  \hfill 
  \begin{subfigure}[b]{\columnwidth}
  \centering
    \includegraphics[width=0.7\linewidth]{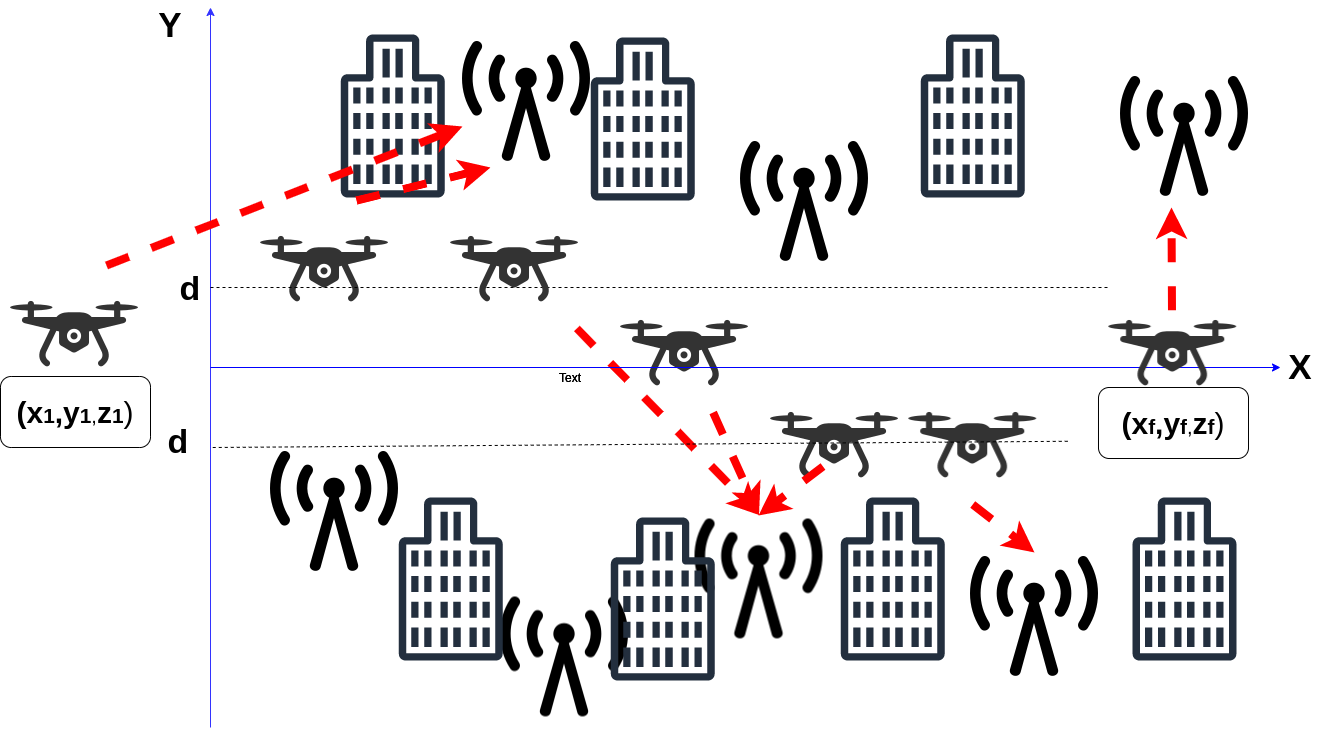}
    \caption{Top view of a UAV connected to the mobile network when it moves closer to the connected BS through the path in an urban area from position  ($x_1,y_{1}$)  to the final position ($(x_f,y_f)$). In red shows which BS the UAV is connected at each point of the trajectory. }
    \label{fig:nonstraight}
  \end{subfigure}
   \hspace*{\fill}%
  \caption{UAV vertical and horizontal movement assumed in this paper.}\label{fig:movements-sim}
\end{figure}

Usually, \acp{uav} are allowed to fly in a height range defined by safety regulation,  with the minimum allowed height denoted as $Z_{min}$, and the maximum allowed height as $Z_{max}$. We assume that the \ac{uav} starts at $Z_{min}$. % is in order to avoid any unnecessary movement, what would increase the \acp{uav} energy consumption. 
Figure \ref{fig:movements-sim} shows \acp{uav} horizontal movement and vertical movement in the generated environment analyses. %We did not illustrate the horizontal path for the experiment because it was a straight line.
Figure \ref{fig:flying-sim} illustrates the possible path of the \ac{uav}, where $d$ is the maximum distance the \ac{uav} can move up or down in each step. It is an representation of a limitation of how much a \ac{uav} can move realistic up or down and horizontally in a time-step.

Our main objective is to optimise $z$ coordinate at each step, in order to improve the  \ac{qos} experienced by the \ac{uav}. The metric used to represent the \ac{qos} is the spectrum efficiency $SE$.

%downlink throughput. The problem assumes the downlink throughput as we use real-world data, and in the data collection, this was the parameter measured. Therefore, the solution could be modified to work with the uplink throughput without modifications to the actual proposed method, this is out of scope of this paper but it may be investigated in future work.  Throughput  is first calculated as symbols per second, where these symbols are defined by their modulation \cite{miao2016fundamentals}. Depending on how many bits a symbol can carry, throughput is converted into bits per second (bps). Thus, throughput can be defined as $T = sM$, where $s$ is the number of transmitted symbols per second, and $M$ is the number of bits per symbol. The modulation is chosen based on the channel conditions.
%, where if the channel conditions are poor, the selected modulation will transmit fewer bits, and if the channel conditions are good, the modulation will transmit more bits. %The physical layer of the transmitter chooses which modulation to use depending on the \ac{SINR}, \ac{ber}, \ac{bler}, etc. 
%We had no access to the exact modulation the UAV had on the experiment or even how many resource blocks it was defined to the \ac{uav}. However, we had access to the throughput the \ac{uav} archived at each step. 

We formulate the optimisation problem as follows:
\begin{subequations}
\begin{align}
 & \underset{(z_1, ..., z_{f-1}, z_f)}{\text{max}} \quad \sum_{t=1}^{f} SE(t); \\
 \textrm{s.t.}\quad & z_t > Z_{min} \quad  \forall \quad t \\
 & z_t < Z_{max} \quad  \forall \quad  t \\
 & |z_t - z_{t-1}| \le d \quad  \forall \quad  t
\end{align}
\end{subequations}

%Where the objective is to maximise the throughput $(T)$ over the path, considering the constraints of the \ac{uav} be inside the allowed altitudes, and that its height change does not surpass $d$.

%In an \ac{LTE} system, the bandwidth is divided by resource blocks. Each bandwidth will have a specific number of resource blocks; for example, 20 MHz has 100 resource blocks with 168 Symbols per second. If the modulation, which depends on channel conditions, is 64 QAM (that transmits 6 bits per symbol), the throughput would achieve 168x100×6=100.8Mbps,  although 25% of the throughput is used for signalling. Once the system can offer MIMO, the achievable throughput increases depending on its order (MIMO order multiplies the achievable throughput without MIMO).

We assume that the \ac{uav} will have access to the \ac{SINR} measurements from its connection, the spectrum efficiency $SE$ of its actual location, and its height at all steps. \ac{SINR} and spectrum efficiency data is easily obtained by the \ac{uav} from its cellular connection, while the height information is obtained via other \ac{uav} sensors located on the UAV.

\section{Proposed Solution}
\label{sec:ml}

\label{sec:proposed-rl-exp}
% !TEX root = ../../main.tex
%In this section, we detail our proposed algorithm.
 \begin{figure}[t!]
    \centering
	\includegraphics[width=\columnwidth]{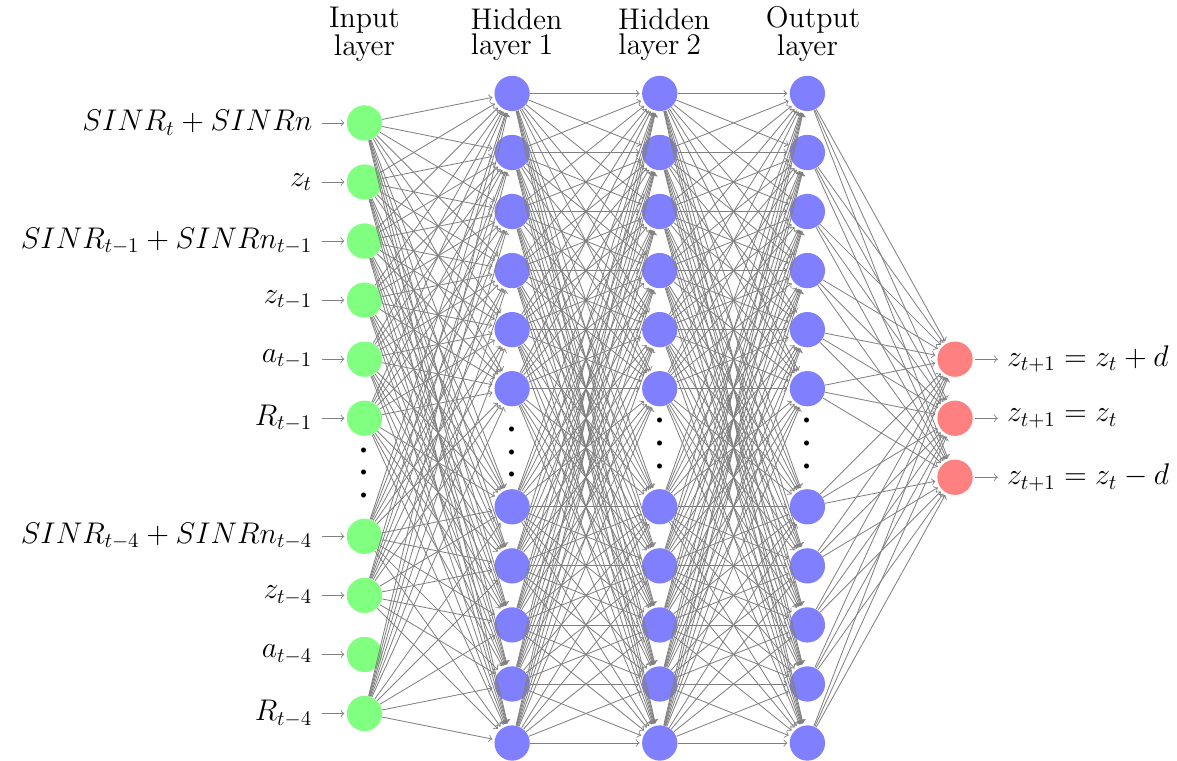}\hfill
	\caption{
Graphical representation of the designed \ac{DL}.
	}
	\label{fig:uav-nn}
	
\end{figure}

To solve the height optimisation problem for a specific position of the \ac{uav} given a particular topology of the \acp{bs} and buildings, one could apply stochastic geometry as in \cite{boris_dic_ant}.
The main issue with this approach is that to represent this problem via stochastic geometry, one has to know the statistical distribution of the features of the environment for each position that the \ac{uav} assumes during flight. This can be computationally expensive to run and the environmental statistics may not be accurate to what the \ac{uav} would find once it is flying in the real world.

\subsection{RL agent definition}
To tackle this issue our solution is  
based on \ac{rl}. 
%\ac{rl} is one of the three main \ac{ML} paradigms and is useful when the problem to be solved does not have an exact label. 
In particular, we apply \ac{dql} as it does not require a predefined model of the environment, since it learns by interacting with the environment in an online manner. %In our implementation, we evaluated the proposed model using a commercial off-the-shelf computer with Intel Core i7-6820 HK processor and GeForce GTX 1070 Mobile.
The agent of our model is the \ac{uav}, as it is the one taking the action of changing the height. Bellow we define the other main components of our model.

\subsubsection*{State Space $S$}
$S$ is all the possible values of the state, and $s$ is the individual single value of the state. We just considered in the state space values a normal \ac{ue} would have from the network and measurements of sensors that a \ac{uav} should have to have a safe fly. Follow the components of $S$:
\begin{itemize}
    \item Height $z$: which is obtained by \ac{uav} sensors and it is relevant for \ac{uav}'s decision-making process, as in order to know whether to move next and stay within the hard limits.
    \item Received $SINR$ and the strongest 5 \ac{SINR} of  neighbours \ac{bs}s-$SINRn$: which is obtained by \ac{ue} sensors to perform the measurement reports. This value impacts the \ac{ue} \ac{qos} that is what we intent to maximise.
    %\item Throughput $T$: which is the actual throughput of the model. This is what we want to maximise in this experiment; it is how we evaluate if the \ac{qos} is improving.
    %\item \ac{SINR} of neighbours \ac{bs}s - $SINRn$: which is obtained from \ac{ue} sensors to report measurament reports.
    \item 4 last $z$, $SINR$, $SINRn$, $a$, $r$: which will be stored from the previous steps. In order to achieve better optimisation, we extended the state with the four previous $z$, $SINR$, action $a$ and reward $r$, following the lines of the original \ac{dql} implementation \cite{sutton2018reinforcement}, as well its implementation in \ac{uav} connectivity \cite{reqiba}.
\end{itemize}

The agent has as input at each time step $s_t$, where t represents the time step the follow state. 
\\
$s = \{SINR_{t}, z_{t}, SINR_{t-1}, z_{t-1},a_{t-1},r_{t-1},SINRn_{t-1},\\SINR_{t-2}, z_{t-2}, a_{t-2},r_{t-2},SINRn_{t-2}, SINR_{t-3}, z_{t-3},\\a_{t-3}, r_{t-3},SINRn_{t-3},SINR_{t-4}, z_{t-4}, a_{t-4},r_{t-4},\\SINRn_{t-4}\}$.

%The main adaptation we made from the proposed solution presented in the experiment approach to the one in the simulation approach is that we use as input for the model the \ac{SINR} of neighbours \ac{bs}s - $SINRn$. We included this information to the model as a normal \ac{ue} has access to it when doing the measurement report. % and it showed better performance on the simulation. With this, we expected to have  more information of the environment and represent it more precisely in challenging deployments. We do not use the neighbours' data as input for the experiment because we did not have access to this information. 
\subsubsection*{Action Space $A$}

 The action is the adjustment of the \ac{uav} height. An action $a \in \{-d,0,+d\}$ will be taken at the end of each time-step, where:
\begin{itemize}
    \item[] $a = -d \Rightarrow z_{t+1} = z_t - d$
    \item[] $a = 0 \Rightarrow z_{t+1} = z_t$
    \item[] $a = d \Rightarrow z_{t+1} = z_t + d$.
\end{itemize}

\subsubsection*{Reward $R$}

As the primary goal of our approach is to improve the \ac{uav} \ac{qos} during flight, our reward at each time step $r_t$ is defined as the spectrum efficiency achieved after the action at point $x_{t+1}$ at height $z_{t+1}$ in the experiment. %For the simulation, we used the spectrum efficiency as it was available in the simulator and is directed related to the \ac{qos} of the \ac{uav} as the throughput. The extension of the simulator is not in the scope of this work.

Our model has three hidden layers, with 200 neurons in each. Figure \ref{fig:uav-nn} illustrates the graphical representation of the proposed \ac{DL}. %It uses Adam as an optimiser, the loss function of sparse categorical cross-entropy, and 200 epochs for the replay training. We arrived at these values after experimenting with those parameters.

In our solution, we use epsilon greedy approach, this is an strategy to balance exploration and exploitation in \ac{rl} algorithms. We selected the initial $\epsilon=1$, where we select an action at random, and we decrease it at every step of the training process until it reaches 0.05, which in our experiment took 30 steps to reach.

\subsection{Hyper-parameters}
We needed to perform a deep investigation to choose the hyper-parameters and design the model. 
We changed several of the hyper-parameters and inputs of the model until finding the proposed one.
These parameters were experimentally selected among a number of model variations in which the number of layers, number of neurons per layer, activation function, number of epochs, regularisation, and the inputs were varied.  As we apply experience replay, the epochs are how many times the model is trained with the mini-batch at each time-step. 
Depending on the complexity of the \ac{dql} network (for example number of input features, number and size of layers), the training can be performed in a few steps, or require thousands or larger number of steps. However, for the \ac{uav} height adaptation scenario it is imperative to have as few training steps as possible, so that the model can learn to optimise height quickly in any new city environment is applied in. %The value of epsilon decay in our evaluation is small compared to other \ac{dql} applications, which makes it learn quickly. %; it is because our model is trained in each new environment.
In order to have a fast adaptation in a new environment, the model needs to adapt its weights quickly to not interfere with the UAV performance at the end of the path. For this evaluation, the value of $\epsilon$ and $\epsilon$-decay are 1 and 0.9 respectively, effectively meaning that the proposed model trains in 30 steps. We apply replay memory as an strategy to accelerate the learning process, where at each step the model trains with the mini batch for the number of epochs.

\subsection{RL algorithm for UAV height optimisation}

The pseudo-code of the \ac{rl} algorithm to optimise $z_t$ is shown in Algorithm \ref{algori}. %It includes the greedy approach for exploring the new environment and the movement of the UAV.
Some parameters must be chosen and passed as input to the code to run the algorithm. 
They are the $Z_{min}$ and $Z_{max}$, the minimum and maximum allowed height that the \ac{uav} could fly. $X$ and $Y$ are the vectors with the horizontal coordinates the \ac{uav} should acquire during its movement. Where $X = {x_1,x_2,..,x_f}$ and $Y = {y_1,y_2,..,y_f}$, as the horizontal path is predefined.
$\epsilon$, $\epsilon_{Decay}$ and $\epsilon_{min}$ are needed to apply the $\epsilon$-greedy approach. The input $\epsilon$ is the starting value for $\epsilon$, and $\epsilon_{Decay}$ is a value that will multiply $\epsilon$ and reduce its value at each interaction until $\epsilon_{min}$. %The $\epsilon$ decrease value is described in lines 22 to 24.

$\beta$ and $\beta_{min}$ are, respectively, the batch array and the minimum size of the batch needed to apply memory replay. While using memory replay, the number of $epochs$ to train the model and the $Discount$ factor to calculate the new Q value ($newQ$) is required. Finally, $var$ is an integer that indicates how often the target model should be updated. The expected output of this algorithm is the \ac{uav} next height in the next step.

The first step of the proposed RL-based algorithm for UAV height optimisation is to initiate the DQN model and the target DQN model, lines 1 and 2, respectively. Then we initialise the \ac{uav} coordinates in line 3 and initialise variable $t$, which refers to the timestep the \ac{uav} is during the each step. The while statement in line 5 is 
the overall while loop that represents the full flight path of the UAV, and has as many steps as that set of $X$ and $Y$.

Inside the step loop, it is needed to update $t$ and collect the current value of $SINR$ and $SINRns$. % needed to define the state $s_t$.
Then, we update the state value in $s_t$.
After that, we randomly select a number, $randomNum$, and compare its value to $\epsilon$ in line 10. This step is necessary to evaluate the comparison of the $\epsilon$-greedy approach. In line 10, we also check $t$ value to be at least 4, as we need the state information values from the last 4 states for the input of the model. 
If the condition is satisfied, which means $randomNum>\epsilon$ and $t>4$, we use the DQN model to predict the best action $a_t$. If the condition is not satisfied, we randomly choose the action $a_t$.
Once the action $a_t$ is defined, we execute it in line 16, ie modify the \ac{uav} height, moving the \ac{uav} up or down if it does not go above the permitted flight boundaries ($Z_{min}$ and $Z_{max}$). %If the action is to maintain the same height, we do not need to apply any changes to $z_t$.
Then, the \ac{uav} also moves based on the sets $X$ and $Y$ its horizontal coordinates to the next position in line 17.
We obtain the reward which represents the quality of our selected action, and is later used to update the learning process. The reward $r_t$ is equal to the measure throughout after executing the action, as shown in line 18.
The $\epsilon$ decrease value is then performed  in lines 19 to 21. The $\epsilon$ decrease is needed to decrease the amount of of random actions we perform once the model is being trained.

\begin{algorithm}
\caption{RL-based algorithm for UAV height optimisation}%\label{euclid}
  \textbf{Input: $\beta$; $\beta_{min}$; \normalfont{\{//Batch parameters\}}\\
  $\epsilon$; $\epsilon_{Decay}$;$\epsilon_{min}$; \normalfont{\{//$\epsilon$-greedy parameters\}}\\
  $d$; $Z_{min}$; $Z_{max}$;$x$;$y$;\normalfont{\{//Coordinates parameters\}}\\
  $epochs$;$var$;$Discount$;\normalfont{\{//Replay memory parameters\}}}\\
\begin{algorithmic}[1]
\STATE \textit{$DQNModel \gets InitialiseDQLModel()$}
\STATE \textit{$targetDQNModel \gets InitialiseDQLModel()$}
\STATE \textit{$(x_t,y_t,z_t) \gets (x_1,y_1,Z_{min})$}
\STATE \textit{$t \gets 0$}
   \WHILE{\textit{ $(x_t,y_t) \neq (x_f,y_f)$}} 
         \STATE \textit{$ t\gets t+1$}
        \STATE \textit{$ SINR_{t} \gets uav.GetCurrentSINR(x_t,y_t,z_t)$}
        \STATE \textit{$ SINRns_{t} \gets Get.SINR.Neighbours(x_t,y_t,z_t)$}
         \STATE \textit{$s_t \gets SINR_{t}, SINRns_{t}, z_t,s_{t}$}
         \STATE $\textit{randomNum} \gets RandomNumber(0-1)$
        \IF{$\textit{randomNum} > \epsilon$ \textit{and} $t > 4$ } 
        \STATE $a_t \gets \textit{DQNModel.predictMaxValue}(s_t)$
        \ELSE
            \STATE $a_t \gets \textit{Random}(+d,0,-d)$
        \ENDIF  \\
        \STATE uav.takeSelectedHeightAction(d,a, $Z_{min}$, $Z_{max}$)
         \STATE \textit{$(x_t,y_t) \gets (x_{t+1},y_{t+1})$}
         \STATE \textit{$r_t \gets uav.GetQoS(x_t,y_t,z_t) $}
         \IF{$\epsilon$ > $\epsilon_{min}$}
         \STATE \textit{$ \epsilon \gets \epsilon * \epsilon_{Decay}$}
         \ENDIF
         \STATE \textit{$s_{t+1} \gets a_{t},r_t,s_{t}.removeState(s_{-4t})$}
         \STATE \textit{$\beta \gets StoreTransition(\beta,s_t,a_t,r_t,s_{t+1})$}\\
         \COMMENT{//Applying replay memory}
        \IF{\textit{$length(\beta) > \beta_{min}$}}
             \STATE \textit{$tempBatch \gets BatchSample(\beta)$}
             \FOR{\textit{i in $1:length(\beta_{min})$}}
               \STATE \textit{$CurrentQValue \gets  \textit{DQNModel.predict}(tempBatch.state)$}
               \STATE \textit{$FutureQValue \gets  \textit{targetDQNModel.predict}(tempBatch.nextState)$}
               \STATE \textit{$maxFutureQ \gets  \textit{FutureQValue.maxValue(i)}$}
               \STATE \textit{$newQ(i) \gets  \textit{tempBatch.reward}+Discount*maxFutureQ$}
               \STATE \textit{$newQTable \gets CurrentQValue(tempBatch.a(i))$}
               \STATE \textit{$newQTable \gets  \textit{DQNModel.update}(tempBatch.state)$}
            \ENDFOR
            \STATE  \small$DQNModel.train(tempBatch.state,newQtable,epochs)$
        \ENDIF
        \IF[//Updating target model]{\textit{$t \% var $}}
        \STATE \textit{$targetDQNModel \gets setWeights(DQNModel.getWeights())$}
        \ENDIF
    \ENDWHILE

\end{algorithmic}
	\label{algori}
\end{algorithm}

After decreasing the value of $\epsilon$, we then save the new state $s_{t+1}$ with the action $a_t$, reward $r_t$ and the state $s_t$, in order to apply the replay memory later.
Therefore, we need to discard the most old values from the previous 4, that refer to the $_{4t}$ timestep, so $s_{t+1}$ has only its last 4 timesteps.
Once we have the values of $a_t,r_t,s_t$ and $s_{t+1}$, we can save them in the batch $\beta$, which will record the last values in order to train the model later using them. % and accelerating its convergence.

To apply replay memory, the batch $\beta$ needs to have a minimum size that is determined before the algorithm starts by $\beta_{min}$. In line 24, we check if this condition is satisfied. If it is not satisfied, we cannot yet apply replay memory. If it is satisfied, a batch sample of size $\beta{min}$ is taken from $\beta$ and saved in the variable $tempBatch$, as illustrated in line 25. For each value in $tempBatch$ in the loop that starts in line 26, we keep in the variable $CurrentQValue$ the update of the Q value made by the actual  DQN model in line 27. Then, update the Q value for the $tempBatch$ next state with the target model and save in the variable $FutureQValue$ in line 28. In line 29, for each value in $tempBatch$, we store the maximum Q value calculated by the target model in $maxFutureQ$. In order to update the new Q value in line 30, $newQ$, for each value in $tempBatch$, we weight the formula by $Discount$ the actual reward of the saved values with the calculated $maxFutureQ$. In possession of the $newQ$ value and the states from $tempBatch$, we calculate the new Q table in lines 31 and 32, $newQTable$, with the values of the chosen actions updated. Then we train the model with the $tempBatch$, the $newQTable$ a number of $epochs$ defined in the input.
We then update, or do not update, the target model in line 36 to 38, and come back to the beginning of the loop.
The target DQN model increases stability during the replay memory implementation, as the target network only updates its weights at each $var$ step. 

The code where we apply Algorithm \ref{algori} is available to the community in our public GitHub\footnote{https://github.com/Erikagpf/DQN-for-UAV-height-adaptation.}.

\section{Evaluation}
\label{sec:evaluation}
% !TEX root = ../../main.tex
We evaluate how our \ac{rl} approach can adapt the \ac{uav} heights with a purpose to optimise the \ac{uav}'s \ac{qos}. 
The main points that we want to evaluate in this section are how the BS density and building densities influence the optimal height of a connected UAV. 

We investigate the BS density influence to the \ac{uav} height as it can influence the interference suffered on the \ac{uav}. Furthermore, as the BSs can be of different heights, the density of the building can also influence the \ac{los} between the \ac{uav} and the BSs, which can interfere with the \ac{qos}. As it was never investigated if the building density influences the connected \ac{uav}, we designed an evaluation on the building density variety and if it affects the approaches.

In order to assess each of these factors separately, we divide this section in three parts. First, we introduce the benchmarks used to compare our proposed approach, then we analyse the mean spectrum efficiency by BS density and building density. Finally, we inspect height changes within each approach. We investigate the mean spectrum efficiency as the \ac{qos} metric that needs to be improved, and we show how the approaches behave on the actual height changes.
%We evaluate how our \ac{rl} approach can adapt the \ac{uav} heights with the objective to optimise the total spectrum efficiency.  %In the sequel, we assess each of these locations separately.
We run the same algorithm in 100 different \ac{mc} trials, simulating 100 different cities for each BS and building density. The evaluation always start the model from scratch, so it does not use the trained weights from the last run emulating a new Mc trial. %Nevertheless, we do it so we do not overfit the model to the data and have a fair evaluation of the learning over an unseen scenario.

%The main points that we want to evaluate are how the \ac{bs} density and building densities influence the optimal height of a connected \ac{uav}. In order to assess each of these factors separately, we divide this section in two sections. First, we analyse the spectrum efficiency by \ac{bs} density and building density. Then, we inspect height changes within each approach. 

The hyper-parameters that provided the best results and were used in the evaluation of the proposed approach are illustrated in Table \ref{table:rl-exp}. 

\begin{table}[t]
\centering
\begin{tabular}{|c|c|}
\hline
\rowcolor[HTML]{C0C0C0} 
Parameters & Value \\ \hline
Epoch & 200 \\ \hline
Epsilon & 1 \\ \hline
Epsilon\_decay & 0.9 \\ \hline
Neurons per hidden layer & 200 \\ \hline
Number of hidden layers & 3 \\ \hline
Regularisation after hidden layers & RELU \\ \hline
Output layer & Softplus \\ \hline
Optimisation function & Adam \\ \hline
\end{tabular}
\caption{
Values of hyper-parameters for the proposed DQN model.
	}
	\label{table:rl-exp}
\end{table}

\begin{figure}[h]
  \centering
  \begin{subfigure}[b]{0.8\columnwidth}
    \includegraphics[width=\columnwidth]{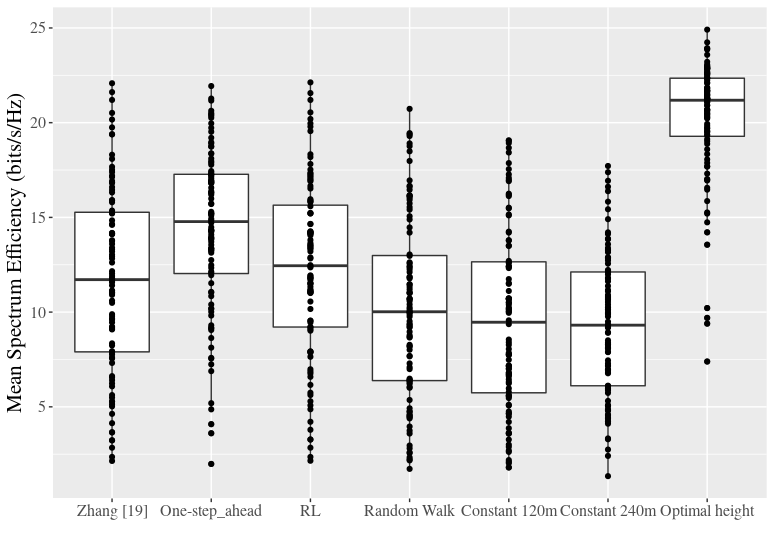}
    \caption{Spectrum efficiency per unit of bandwidth (bits/s/Hz) for low building density}
    \label{fig:rate1}
  \end{subfigure}
  \hfill 
  \begin{subfigure}[b]{0.8\columnwidth}
    \includegraphics[width=\columnwidth]{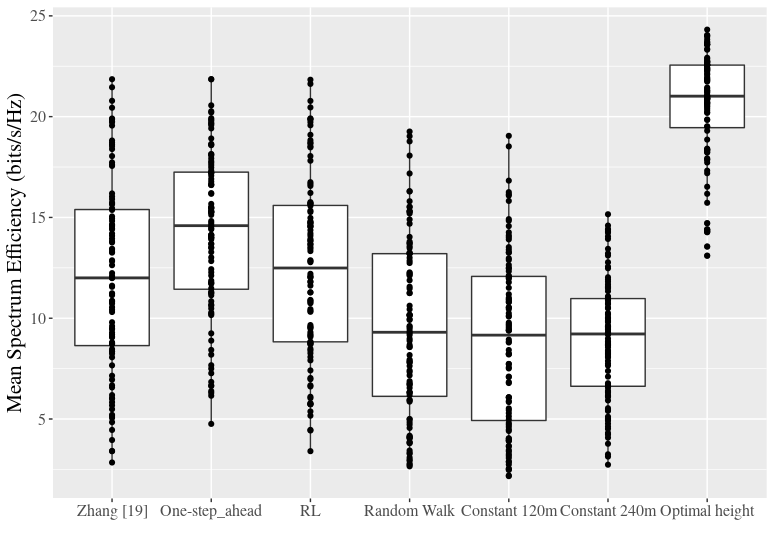}
    \caption{Spectrum efficiency per unit of bandwidth (bits/s/Hz) for medium BS density.}
    \label{fig:rate2}
  \end{subfigure}
    \begin{subfigure}[b]{0.8\columnwidth}
    \includegraphics[width=\columnwidth]{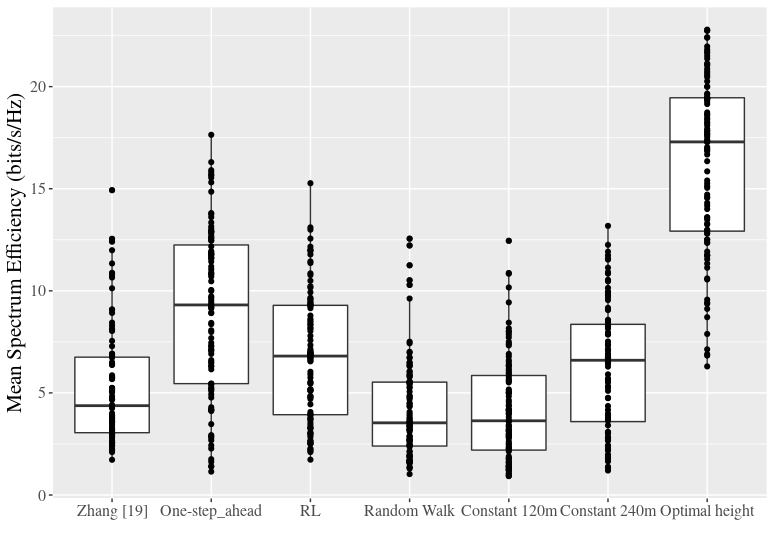}
    \caption{Spectrum efficiency per unit of bandwidth (bits/s/Hz) for high BS density.}
    \label{fig:rate3}
  \end{subfigure}
  \caption{Spectrum efficiency per unit of bandwidth (bits/s/Hz) for 3 different BS densities and medium building density.}\label{fig:rate-box}
\end{figure}

\subsection{Baselines}
\label{sec:baseline}
We choose five different height selection strategies to which we compare performance of our proposed \ac{rl} algorithm. 
For the first one, we use the baseline proposed by Zhang \cite{trajec-mobile}, which suggests that the \ac{uav} maintain the minimum allowed height during its flight. One of the most common approaches to \ac{uav} height selection is to maintain a constant height \cite{nallanathan,rami1,reshape,network-adapt}, but there is no consensus on which height value to choose. To make a fair comparison, we also benchmark our approach against two constant height values. These heights will be the maximum possible height (120 m), and half of the maximum (60 m). When following these fixed height strategies, the \ac{uav} will begin at the minimum height at timestep 1, before increasing its height in each timestep until it reaches the required height, after which it will make no further adjustments.

To confirm that our solution is actually learning based on observed environment information and not acting randomly, we also compare it to a bounded Random walk height selection strategy, in which the \ac{uav} in each timestep randomly selects one of  three actions: increase the height, decrease the height, or keep the current height. It is bounded as all the solutions and cannot fly outside the allowed flight range. 

In order to compare our solution to a more complex baseline, we implement an approach that we call One-step-ahead solution. In the One-step-ahead approach, the \ac{uav} knows whether the maximum \ac{SINR} in the next time step will be found above or below its current height, and will move up or down (in a fixed increment of d = 10 m) depending on this knowledge. To be able to apply the One-step-ahead solution, the \ac{uav} needs previous information about  the environment; this is not feasible in a real-world application, but we include this to assess whether and by how much  such information would improve performance when compared to our \ac{rl} approach.

We also compare our \ac{rl} solution with  one based on optimal height at each time step as obtained from the real-world dataset. In this approach, it is assumed that the \ac{uav} is able to move to any height in the next timestep, without restrictions of $d$. This represents the ideal-case performance which would not be possible in a real-world \ac{uav} application.

Bellow are benchmark approaches:
\begin{itemize}
    \item \textbf{Zhang \cite{trajec-mobile}}: this benchmark proposes that the \ac{uav} should maintain the minimal allowed height at all times.
    \item \textbf{Constant at  60 m}: this benchmark starts at the minimal height, like all others, and then moves up at every step until it achieves 60 m height. After achieving 60 m, the \ac{uav} should not move up or down.
    \item \textbf{Constant at  120 m}: this benchmark starts at the minimal height, like all others, and then moves up at every step until it achieves 120 m height. After achieving 120 m, the \ac{uav} should not move up or down.
    \item \textbf{Random walk}: this benchmark chooses its action randomly at each step.
    \item \textbf{Optimal height}: this is a reference benchmark of the maximum possible \ac{qos} values. In this approach, the \ac{uav} does not have any limitations on the maximum height change from step to step and also know which height has the maximum \ac{qos}. 
    \item \textbf{One-step-ahead}: this benchmark follows the optimal height next position to decide its next action. If in the next step the optimal height is above the actual height of the \ac{uav}, the chosen action will be to move up. If in the next step the optimal height is bellow the actual height of the \ac{uav}, the chosen action will be to move down. In case the optimal height in the next step is the same as the actual height, the \ac{uav} should not move.
\end{itemize}

\subsection{Spectrum efficiency}
In this section we analyse the mean of spectrum efficiency per unit of bandwidth, that is a mean of the spectrum efficiency over an entire episode, for varying \ac{bs} densities and building densities. We inspect the spectrum efficiency as this is the parameter that we wish to optimise.%and how the height and spectrum efficiency changed through one single episode, in section \ref{sub:heightbs}. %Here I get each run, make the mean, and generate the graph.

\subsubsection{Varying BS densities} \label{sec:spectralbs}

To demonstrate how the \ac{rl} solution can have its performance affected by different \ac{bs} densities, %and evaluate when it turns necessary or not, 
we study in detail three different \ac{bs} densities $(1,2.5,5)/km^2$, denoted as low, medium and high, as illustrated in Figure \ref{fig:rate-box}. %We choose these values as an urban scenario is usually represented by 5 \acp{bs} per $km^2$, and rural is denoted by 1 \acp{bs} per $km^2$ \cite{}.

Figure \ref{fig:rate1} shows the mean spectrum efficiency per approach. As expected, the optimal height provides much better spectrum efficiency, achieving median of 23 bits/s/Hz. This happens because it does not have any movement restriction, being able to move any distance from step to step. For low \ac{bs} density, One-step-ahead, Zhang \cite{trajec-mobile} and the proposed \ac{rl} approach perform similarly, with all archiving median of 20 bits/s/Hz. The constant height at 240 m is the approach with the worse spectrum efficiency, with 15 bits/s/Hz, showing that high heights for low \ac{bs} density do not perform as good as other approaches do. Constant at 120 m performed slightly worse than the Random walk approach, with median of 18.5 bits/s/Hz and Random walk approach with 19 bits/s/Hz. It is interesting to note that the approaches do not vary much its mean spectrum efficiency, and all have a relatively small first and third quartile of around 2 bits/s/Hz, with exception of Constant at 240 m with 4 bits/s/Hz. 

Figure \ref{fig:rate2} shows that our \ac{rl} approach performs better, 4\%, then Zhang \cite{trajec-mobile} for medium \ac{bs} density, and 26\% better than Constant at 120 m, Constant at 240 m and Random walk. It indicates that maintaining higher heights at all times provides worse spectrum efficiency for the medium \ac{bs} and building densities when compared to the proposed \ac{rl} approach that adapts the height dynamically to the environment. One-step-ahead showed the best performance compared to the approaches that could only move "d", achieving 14.5 bits/s/Hz, showing that for medium \ac{bs} density having previous knowledge of the radio characteristics of the environment can improve the \ac{uav} \ac{qos}. 
%However, as at 240 m the spectrum efficiency is higher in 18\% than at 120 m, shows that this is not a rule and that the coverage in the sky is a complicated matter, as the \ac{ue} might be connected to a side lobe when it is at different heights and distances to the serving \ac{bs}.

When investigating the high \ac{bs} density in Figure \ref{fig:rate3}, Constant at 120 m and Random walk are the worst solutions achieving 3.5 bits/s/Hz, with the Zhang \cite{trajec-mobile} being slightly better than them, showing that maintaining the lowest altitude for all topologies is not the best approach. The proposed \ac{rl} approach shows performance comparable to Constant at 240 m, with 4\% better performance. Therefore, its third quartile is higher, which means that the \ac{rl} performed better in more runs. The One-step-ahead approach showed the best performance with its median achieving 9 bits/s/Hz, showing the previous knowledge of the environment can improve \ac{uav}s \ac{qos}. However, it is unrealistic to expect to have this knowledge for each set of coordinates in the environment.

%For both, medium and high BS densities, the optimal height difference to the other approaches increased considerable when compared to the minimum height.
\iffalse
 \begin{figure}[t]
    \centering
	\includegraphics[width=0.8\columnwidth]{sections/images/RateBSno.png}\hfill

	\caption{Average Spectrum efficiency per unit of bandwidth (bits/s/Hz) per BS density (low, medium, and  high), and medium building density.
	}
	\label{fig:overalrate}
\end{figure}

We have used Figure \ref{fig:rate-box} to analyse how each approach compares to each other, now Figure \ref{fig:overalrate} shows the data in a different format to easier show how the \ac{bs} density impacts all the approaches performance.
\fi
 When analysing a macro view between the different densities, Figure \ref{fig:rate-box} shows that the general mean spectrum efficiency for low \ac{bs} density is much better than for medium and high \ac{bs} density, with solutions archiving near 20 bits/s/Hz. We can also analyse that One-step-ahead and the proposed \ac{rl} solution are always the best approaches for all densities, showing that an intelligent and adaptable decision can provide a good \ac{qos} for all densities. Moreover, the proposed \ac{rl} solution can adapt its response to the environment on the fly without previous knowledge.%, different from One-step-ahead and is consistently the second best approach to all densities, being the same as Zhang \cite{trajec-mobile} only at low BS density.

\subsubsection{Varying building densities}
%In this section we evaluate how the building density impacts the overall spectrum efficiency. To the best of our knowledge, this is the first work to evaluate how the building density affects the \ac{uav}-BS link.

\begin{figure}[t]
  \centering
  \begin{subfigure}[b]{0.8\columnwidth}
    \includegraphics[width=\columnwidth]{sections/images/bs4-plot.png}
    \caption{Spectrum efficiency per unit of bandwidth (bits/s/Hz) for low building density}
    \label{fig:rate4}
  \end{subfigure}
  \hfill 
  \begin{subfigure}[b]{0.8\columnwidth}
    \includegraphics[width=\columnwidth]{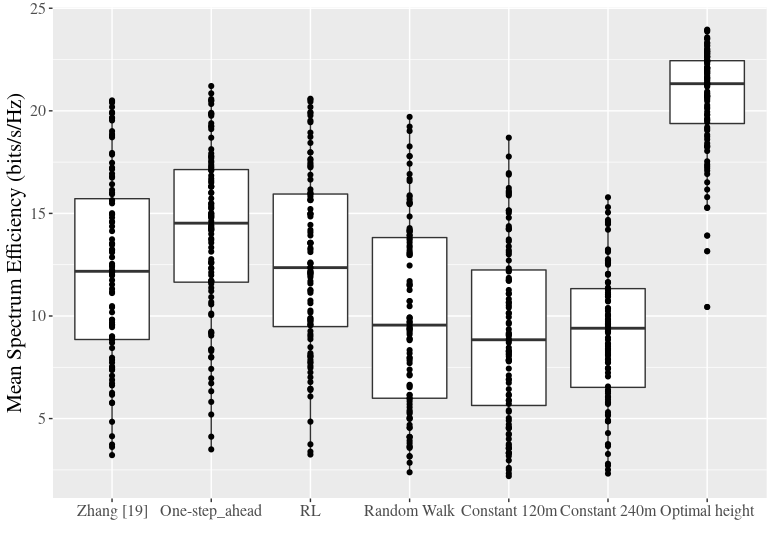}
    \caption{Spectrum efficiency per unit of bandwidth (bits/s/Hz) for high building density.}
    \label{fig:rate5}
  \end{subfigure}
  \caption{Spectrum efficiency per unit of bandwidth (bits/s/Hz) for different building densities with medium BS density.}\label{fig:ratebuild-sim}
\end{figure}

Figure \ref{fig:ratebuild-sim} illustrates the spectrum efficiency for low and high building density.
In Figure \ref{fig:rate4}, the One-step-ahead provides the best approach achieving median of 15 bits/s/Hz, and the proposed \ac{rl} approach is the second best with 12.5 bits/s/Hz. We can observe that Zhang \cite{trajec-mobile} approach achieves 11.7 bits/s/Hz, that is 6\% worse than the proposed \ac{rl} solution. The Constant at 240 m performs as well as the Constant at 120 m, and both are worse than all other solutions, which show a deterioration for those heights, implying that the \ac{uav} would be most of the time in a poor coverage area. Random walk approach performed slightly better then the higher constant approaches, showing that the Random walk movement of the \ac{uav} is comparable to maintaining high constant values.

Figure \ref{fig:rate5} illustrates the mean spectrum efficiency for high building density. It shows a similar pattern when compared to the low building density, with One-step-ahead being the best approach and the proposed RL solution being slightly better, 2\%, than Zhang \cite{trajec-mobile}.
We can conclude that since it has no impact, it is providing an indication that building density is not a factor that needs to be taken account when determining \ac{uav}'s height. It shows that that same approach should work in density urban areas and rural ones.
\iffalse
\begin{figure}[t]
    \centering
	\includegraphics[width=0.8\columnwidth]{sections/images/RateBuildno.png}\hfill
	\caption{Spectrum efficiency per unit of bandwidth (bits/Hz) per building density ( low, medium, and high), and medium BS density.
	}
	\label{fig:overalratebuild}
\end{figure}
Figure \ref{fig:overalratebuild} is a macro view of the performance of the solutions over the building densities in order to show how the building density impacts all the approaches performance.
The medium density is the same as in Figure \ref{fig:overalrate}, but we leave it in Figure \ref{fig:overalratebuild} as it makes it easier to analyse the overall performance.
\fi
As an overall performance between the three different densities, we discovered that the difference in the building density when the UAV is flying above the buildings did not influence the mean spectrum efficiency as the approaches performed similar in all the distributions.

%when the density is low, the maximum mean spectrum efficiency is lower, and it increases as the building density increases. We sugest that it is due to the interference from \ac{bs}s degrade the spectrum efficiency. With fewer buildings to block interference, the mean spectrum efficiency decreases, and once there are building blocking the interference signal, the mean spectrum efficiency can increase. We can also notice that the proposed \ac{rl} solution can adapt to the changes of the environment, and even if in a favourable environment or an environment with more interference, it can provide one of the bests \ac{qos}.

\subsection{Height variation} \label{sub:heightbs}
%As we are evaluating how to choose the most appropriated height for the \ac{uav} to increase rate,
While in the previous section we focus our analyses on the spectrum efficiency of each approach, in this section we inspect in more detail underlying height variations that achieve the discussed performance.

\begin{figure}[t]
  \centering
  \begin{subfigure}[b]{0.8\columnwidth}
    \includegraphics[width=\columnwidth]{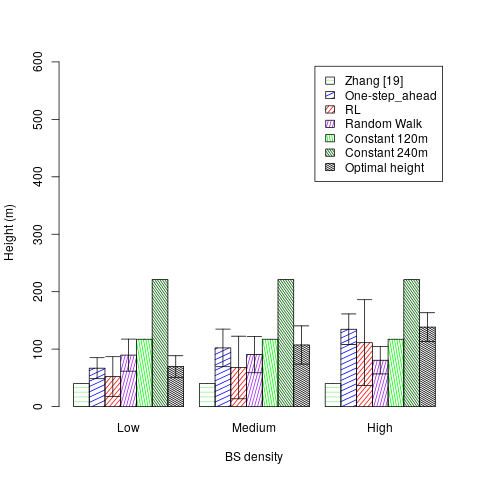}
    \caption{Mean height per BS density.}
    \label{fig:heightdensbs}
  \end{subfigure}
  \hfill 
  \begin{subfigure}[b]{0.8\columnwidth}
    \includegraphics[width=\columnwidth]{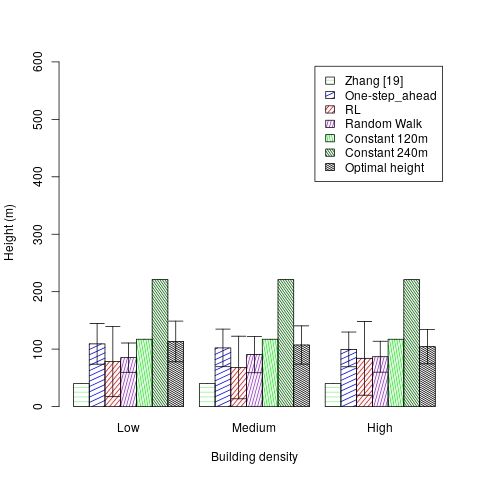}
    \caption{Height per building density.}
    \label{fig:heightdensbuild}
  \end{subfigure}
  \caption{Height analyses for varying BS and building densities.}\label{fig:heightdens}
\end{figure}

To make a more detailed investigation over the 100 \ac{mc} trials, Figure \ref{fig:heightdens} illustrates the mean of the heights for different \ac{bs} and building densities. The constant approaches have no variance on the height after they achieve their constant heights. In Figure \ref{fig:heightdensbs}, the average height of the optimal height approach varies with the \ac{bs} density, being lower for low \ac{bs} density, and higher for high \ac{bs} density. As we can notice, the intelligent approaches, One-step-ahead and the proposed \ac{rl} solution, adapt their altitude to the one that better serves the \ac{bs} distribution, also increasing its heights when the \ac{bs} density increases. The Random walk approach, as it does not consider any information of the environment, it also maintains, in average, the same height in all cases.

When we analyse in Figure \ref{fig:heightdensbuild} the height adaptation by the building density, the optimal height is not related with the density. The approaches does not change its mean height considerably during the different building densities. The \ac{rl} approach varies from 68 m in medium building densities, to 83 m in high building densities. 

Observing behaviours for both \ac{bs} and building densities, we conclude that \ac{rl} is a competent approach to solve \ac{uav} height optimisation. As we can see in Figure \ref{fig:heightdens}, the \ac{rl} solution demonstrated to be learning the best height, resulting in a spectral efficiency improvement.
We can also conclude that the \ac{rl} approach does not make changes on its height at all steps, making intelligent changes when needed and avoiding spending extra energy to move its height at all steps.
\section{Real-world Data Evaluation}
\label{sec:evaluation-exp}
In this section, we evaluate how our \ac{rl} approach can adapt the \ac{uav} heights with the objective to optimise the total throughput.  %In the sequel, we assess each of these locations separately.
We first introduce the experimental measurement data-set and then provide a detailed evaluation of the proposed solution using the real-world dataset.

\subsection{Experimental measurement Dataset}
\label{sec:data-exp}
% !TEX root = ../../main.tex

To evaluate the proposed height adaptation solution, we also use the real-world measurements obtained by a \ac{uav} connected to a two-tier cellular network in two different areas of Dublin city’s Smart Docklands, which includes massive \ac{mimo} macro cells and \ac{mimo} small cells. Below, we recap the details of the experiment relevant for our evaluation, while full details of measurements are presented in \cite{measurament}. 

\begin{figure}
\centering
	\includegraphics[width=0.8\columnwidth]{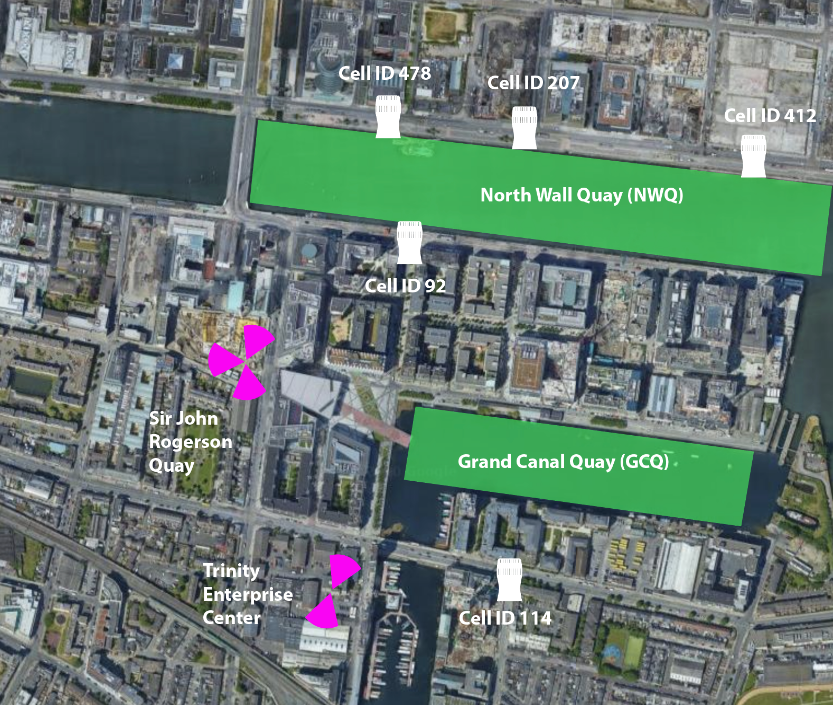}
	\caption{
Testbed area from the top. Macro cells are labelled in purple, and small cells are denoted with white icons—the measurement area where the UAV flew is marked in green.
	}
	\label{fig:testbed}
\end{figure}

The experimental cellular network testbed, in which the measurements were conducted, is shown in Figure \ref{fig:testbed}. Connectivity data was collected in two environments: Grand Canal Quay (GCQ) and North Wall Quay (NWQ), as illustrated in Figure \ref{fig:testbed}. The \ac{uav} flew at a fixed height  back-and-forth in the designated areas. This flight pattern was repeated at 10 meter increments for all heights between 30 and 120 meters (the legal flight ceiling in Dublin).

Table \ref{table:experiment-exp} summarises the main characteristics of the experimental environment for NWQ and GCQ. The Table shows: the height of the \ac{bs} antennas; the velocity the \ac{uav} was flying; the total distance the \ac{uav} passed in each height; the quantity of measurements reports in each area, denoted as steps; the size of each step in meters; the variation of heights; the building height variation; and the distance $d$ in each scenario. 
The measurements were reported every 2 seconds most of the time. We also observe that the flight in NWQ resulted in fewer measurements despite being the one where the \ac{uav} flies for a longer distance. 
While a \ac{ue} is performing handover, it does not sense the spectrum; consequently, it does not report any measurement. In the small cell area, the \ac{uav} was performing handovers, which resulted in fewer measurement reports when compared to the macro cell area, where the \ac{uav} did not perform measurement reports.
%result in the interval between some measurements being longer than 2s. 

\begin{table}[t]
\centering
\begin{tabular}{|c|c|c|}
\hline
\rowcolor[HTML]{C0C0C0} 
Variable & NWQ Value & GCQ Value\\ \hline
BS height &  $6.5 m$ & $29 m$ \\ \hline
Speed & 4.2 $m/s$ & 2.6 $m/s$ \\ \hline 
UAV travel distance  & 1160 $m$ & 890 $m$ \\ \hline 
Steps & 161 & 171 \\ \hline 
Horizontal step size & 7.2m & 5.2 \\ \hline 
Allowed UAV height range & $[30-120] m$ & $[20-120] m$ \\ \hline 
Building height variation & $[20-80] m$ & $[20-80] m$ \\ \hline 
d & $10 m$ & $10 m$ \\ \hline 
\end{tabular}
\caption{
Collection environments.
	}
	\label{table:experiment-exp}
\end{table}

In order to use the proposed approach with the available real-world data we had slightly modify the definition of an RL agent. In the real-world data the \ac{qos} information available is the throughput, so we used this information instead of the spectrum efficiency in the proposed solution. The real-world dataset also had no information about the sensed neighbours, so we do not include this as input of the model. The remainder or the algorithm is exactly the same as in the generated environment.
The final state space of the adapted solution is: 
$s = \{SINR_{t}, z_{t}, SINR_{t-1}, z_{t-1},a_{t-1},r_{t-1},SINR_{t-2}, z_{t-2},\\a_{t-2},r_{t-2}, SINR_{t-3}, z_{t-3},a_{t-3}, r_{t-3},SINR_{t-4}, z_{t-4},a_{t-4},\\r_{t-4}\}$.
A sample of the used data in illustrated in Table \ref{table:sample-data-uav-changed}.
\begin{table*}[t]
\centering
\begin{tabular}{|c|c|c|c|c|c|c|}
\hline
\rowcolor[HTML]{C0C0C0} 
Height & Step & Latitude & Longitude & SINR Carrier 1 (dB)  & Serving Cell Identity &  Throughput (kbps)\\ \hline
20 & 1 & 53.34342193604 & -6.23032475884304 & 10.3
 &  60 & 46761.22  \\ \hline
30 & 1 & 53.34342193604 & -6.23032475884304 & 20.2 &  60 & 76651.78  \\ \hline
40 & 1 & 53.34342193604  & -6.23032475884304 & 11.1
&  60 & 46797.87 \\ \hline 
50 & 1 & 53.34342193604 & -6.23032475884304 & 6.4
 &  61 &  35082.27\\ \hline 
60 & 1 & 53.34342193604 & -6.23032475884304 & 3.4
&  61 & 29024.27 \\ \hline 
70 & 1 & 53.34342193604 & -6.23032475884304 & 9.9
& 60 & 46738.80 \\ \hline 
\end{tabular}

\caption{
Sample of data from GCQ area used in the evaluation.
	}
	\label{table:sample-data-uav-changed}
\end{table*}

\begin{table}
\centering
%\resizebox{\columnwidth}{!}{%
\begin{tabular}{|c|c|}%c|}
\hline
\rowcolor[HTML]{C0C0C0} 
Approach & Throughput (Mbps)  
\\ \hline
Zhang \cite{trajec-mobile} & 35 
\\ \hline
Constant at  60 m & 30 
\\ \hline
Constant at  120 m & 30  
\\ \hline
Random walk & 32 +- 2  
\\ \hline
One-step-ahead & 35  

\\ \hline
Optimal height &  43   
\\ \hline
RL  &  37 +- 1   
\\ \hline
\end{tabular}
%}
\caption{
Mean throughput (Mbps) over 100 trials. Based on flight data obtained in the NWQ area. 
	}
	
	\label{table:nwq}
\end{table}

\subsection{Evaluation of the proposed RL approach}

%We run the same algorithm 100 times over the data for this evaluation but always start the model from scratch, so it does not use the trained weights from the last run. Nevertheless, we do it so we do not overfit the model to the data and have a fair evaluation of the learning over an unseen scenario.
We evaluate performance of our approach in two different sets of real-world data: data collected in NWQ, with small cell connectivity, Section \ref{sec:nwq}, and data collected in GCQ, with macro cell connectivity, Section \ref{sec:gcq}. % with characteristics as described in Section \ref{sec:data-exp}.
We start the evaluation with the throughput analysis, followed by the analysis of the height adaptation through the path.
We evaluated the model after the training phase in this section. The results shown are related to the last 100 \ac{uav} steps. 

\subsubsection{NWQ analysis}
\label{sec:nwq}

\begin{figure}[t]
  \centering
  \begin{subfigure}[b]{0.8\columnwidth}
    \includegraphics[width=\columnwidth]{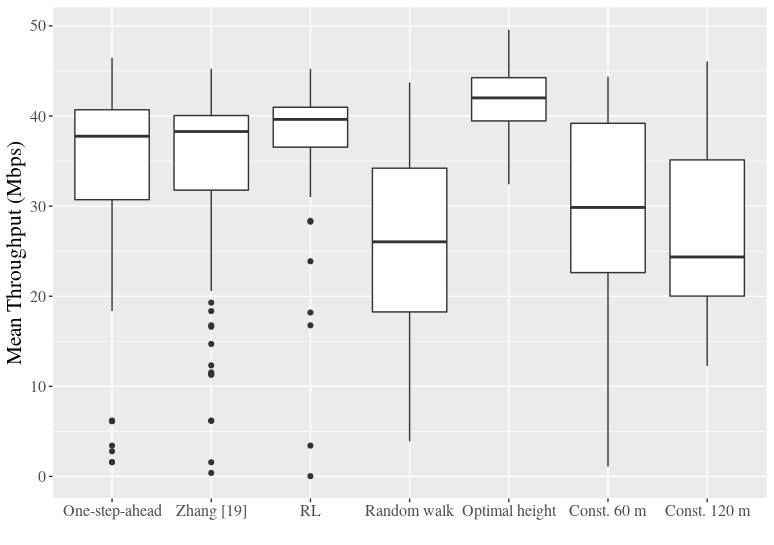}
    \caption{Mean throughput in (Mbps) in NWQ area.}
    \label{fig:nwqrate}
  \end{subfigure}
  \hfill 
  \begin{subfigure}[b]{0.8\columnwidth}
    \includegraphics[width=\columnwidth]{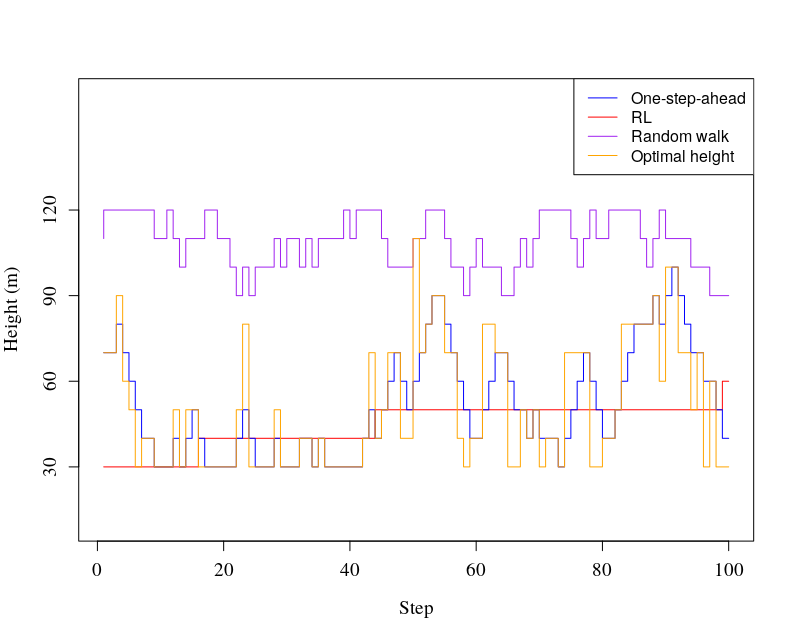}
    \caption{Height adaptations for a \ac{uav} flight over the last 100 steps in NWQ area.}
    \label{fig:nwqheight}
  \end{subfigure}
  \caption{Performance of the benchmarks and the proposed RL approach in the NWQ area.}\label{fig:nwq}
\end{figure}

%In this subsection, we analyse the throughput obtained from height adaptations based on data collected during flights performed in the NWQ area, where the \ac{uav} was connected to different small cells during its flight.  
Table \ref{table:nwq} presents the average throughput of the investigated approaches; for non-deterministic solutions, which means the ones that might change at each run, we present a mean over 100 trials. 
%It also presents the number of height changes each of the solutions performed during the flight.
We inspect the throughput as it is our parameter that we wish to optimise.
By construction, the Optimal height at each step leads to the highest throughput. Therefore, we consider the Optimal height at each timestep to be the one with the highest throughout at that timestep.
Our proposed approach achieves 37 Mbps with a variance of 1 Mbps, which is the highest throughput on the feasible solutions. One-step-ahead achieves 35 Mbps, that is the second highest. %, with 6\% performance drop from our proposed approach.
The approach proposed by Zhang~\cite{trajec-mobile} performs similarly to the One-step-ahead solution with 35 Mbps, with the added benefit of not needing a priory knowledge of the environment. Nonetheless, our proposed \ac{rl} approach provides the best throughput and outperforms Zhang~\cite{trajec-mobile} and the One-step-ahead benchmarks by 6\%, also resulting in lower variation in \ac{uav} heights when compared to the  One-step-ahead approach. It is worth noting that the solutions that maintained large heights, as Constant at  60 and 120 m, do not perform well when compared to those that maintained lower heights. One possible explanation for this is 
that at greater heights a \ac{uav} might have been experiencing increased interference from cells it was not connected. Another possibility is antenna misalignment: as the small cells are designed for ground users, their antennas are directed towards the ground, which means that the aerial \ac{uav} receives signals primarily from antenna side-lobes.

Figure \ref{fig:nwq}  evaluates an %randomly chosen 
example run, different than Table \ref{table:nwq} that evaluates the approaches performance after 100 trials. To generate Table \ref{table:nwq} we needed to calculate the mean over the throughput mean of each run, losing information of the throughput variation through the path. With the analyses of one single run, we can verify how the throughput and height vary through the path.
Figure \ref{fig:nwqrate} presents box plots for the throughput in Mbps for all approaches obtained across the last 100 steps of one example run. Our \ac{rl} approach shows a stable value for the obtained throughput, with its first and third quartile being 36 to 41 Mbps (the box denotes that 50 \% of the data is in this range), respectively, and with median 40 Mbps. On the other hand, one can observe a considerable interquartile range from 18 to 34 Mbps in the throughput for the Random walk approach, as well as for the approaches that maintain the height Constant at  60 m and 120 m. This more significant variance is likely due to the randomness of the Random  walk approach and to the fact that at greater heights of the constant strategies, the coverage from several cells is more unpredictable, as the \ac{uav} may be connecting to the side lobes of different antennas. Approaches as One-step-ahead and Zhang\cite{trajec-mobile} have a bigger interquartile when compared to the proposed \ac{rl} approach, with Zhang~\cite{trajec-mobile} being between 32 to 40 Mbps, One-step-ahead between 31 to 41 Mbps, and the \ac{rl} approach between 36.5 Mbps to 41 Mbps. Although in the One-step-ahead, Zhang \cite{trajec-mobile} and \ac{rl} happens outliers (in the figure represented as the dots outside the box) that means that at some points of the path, the measured throughput was much lower than most of the path. Interestingly, the Optimal height median throughput is only 6\% better than our RL-based approach, despite it unrealistically assuming instant jump from any height to any other height is possible, showing that the proposed method is close to optimal.

Figure \ref{fig:nwqheight} shows how the different adaptive strategies adjusting the \ac{uav} height at different steps in a single sample run for the last 100 steps. We inspect the individual height adaptation to understand how each of the approaches behave in a real path and have an idea of how many adaptation were needed to achieve their respective throughput.
We do not illustrate Zhang \cite{trajec-mobile}, Constant at  60 m and Constant at  120 m because their values are constant. %We illustrated Zhang \cite{trajec-mobile} because it is a benchmark that was proposed 
%We did not present the Constant at  60 and 120 m, as they do not have any change on their heights. 
We can observe that our proposed \ac{rl}-based solution maintains the \ac{uav} height low all the path, with only 3 changes in the UAV height on the last 100 steps. On the other hand, we can see that the Optimal height at each step changes substantially, indicating that even if one knew in advance at which height the optimal connectivity was obtained, the \ac{uav} would not be capable of reaching these heights in every timestep, as the height change from one step to another could be in the order of 90 m. The One-step-ahead approach follows the Optimal height, and also moves constantly trying to achieve the Optimal height approach. In this example, the Random walk approach started the last 100 steps at higher heights and it moved randomly through the steps in a up and down movement, and sometimes, did not move, as expected. 
%, which show an economic use of the \ac{uav}'s energy as it does not keep moving constantly

\subsubsection{GCQ analysis}
\label{sec:gcq}
\begin{table}
\centering
%\resizebox{\columnwidth}{!}{%
\begin{tabular}{|c|c|}%c|}
\hline
\rowcolor[HTML]{C0C0C0} 
Approach   & Throughput
\\ \hline
Zhang \cite{trajec-mobile}  & 68 
\\ \hline
Constant at  60 m  & 41 
\\ \hline
Constant at  120 m   & 41 
\\ \hline
Random walk   & 50 +- 4 
\\ \hline
One-step-ahead   & 68  

\\ \hline
Optimal height  & 83 
\\ \hline
RL    & 70 +-2 
\\ \hline
\end{tabular}
%}
\caption{
Mean throughput (Mbps) over 100 trials. Based on flight data obtained in the GCQ area. 
	}
	
	\label{table:gcq}
\end{table}

Table \ref{table:gcq} shows the average throughput for the GCQ area over 100 trials. Same as in NWQ, we aim to analyse the throughput as it is the variable that we intent to optimise. 
The Random walk approach provided a throughput of 50 Mbps, better then the constant approach at 60 with and 120 m that achieve.
The constant approaches that lead to the \ac{uav} flying at larger heights result in lower throughput compared to all other approaches, obtaining 41 Mbps, which is only 59\% of the throughput achieved by our \ac{rl} approach.
In this scenario, our \ac{rl} solution also performed better than all benchmarks achieving 70 Mbps in average, while the Zhang \cite{trajec-mobile} approach and One-step-ahead being in second, achieving 68 Mbps. The results of the One-step-ahead approach show that having a priori knowledge of the environment is sometimes not enough to provide the best throughput. 
As a reference, the Optimal height achieved around 19\% better throughput than the proposed \ac{rl} approach, which showed to be considerate more than in NWQ area. One explanation of the difference in the distance between the Optimal height and the other methods is due to the fact that the optimal approach changed more drastically its height through the path, making it impossible for any other approach to achieve closer to the same throughput as they were limited by "d".

\begin{figure}[t]
  \centering
  \begin{subfigure}[b]{0.8\columnwidth}
    \includegraphics[width=\columnwidth]{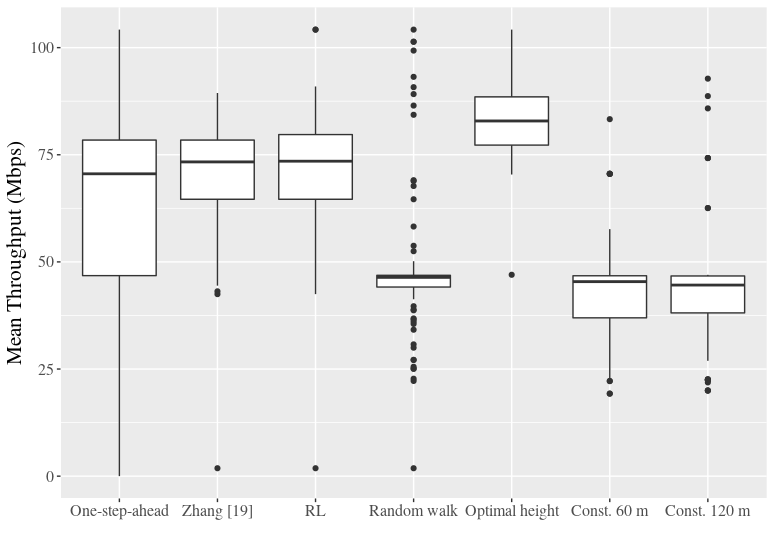}
    \caption{Mean throughput in (Mbps) in GCQ area.}
    \label{fig:gcqrate}
  \end{subfigure}
  \hfill 
  \begin{subfigure}[b]{0.8\columnwidth}
    \includegraphics[width=\columnwidth]{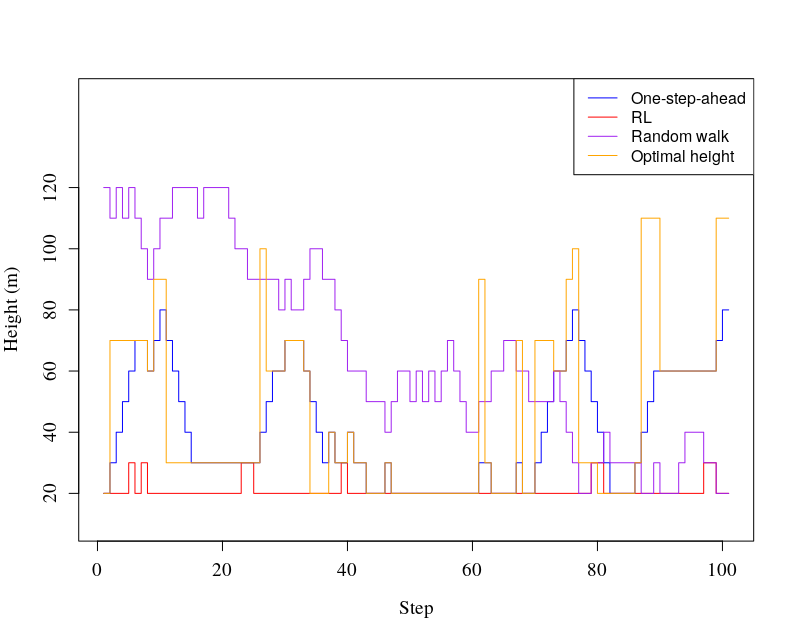}
    \caption{Height adaptations for a \ac{uav} flight over the last 100 steps in GCQ area.}
    \label{fig:gcqheight}
  \end{subfigure}
  \caption{Performance of the benchmarks and the proposed RL approach in the GCQ area.}\label{fig:gcq}
\end{figure}

As in the NWQ area, Figure \ref{fig:gcq}  evaluates an %randomly chosen 
example run, different than Table \ref{table:gcq} that evaluates the approaches performance after 100 trials. %To generate Table \ref{table:gcq} we needed to calculate the mean over the throughput mean of each run, losing information of the throughput variation through the path. With the analyses of one single run, we can verify how the throughput, height and learning vary through the path.
In Figure \ref{fig:gcqrate}, we investigate the stability of each of the approaches, with the box plot representing throughput across last 100 steps. Both, \ac{rl} and Zhang \cite{trajec-mobile} approaches, achieve median throughput of 74 Mbps, as well as exhibiting low variance. Both achieve the lower quartile at 65 Mbps, but at the third quartile, the RL proposed approach provides 2 Mbps more than Zhang \cite{trajec-mobile}, meaning that it provided better throughput for some time in the path. % with the first quartile, being 65 MHz for the first quartile for both approaches, and for the third our approach provides  80 MHz.
This behaviour is similar to the one in the NWQ area, although the throughput results for the other baseline approaches are significantly different. In particular, the approaches that keep the \ac{uav} height Constant at 60 and 120 m show lower variance than for the data set obtained in the NWQ area. Possibly this difference is because the \ac{uav} connects to only one macro \ac{bs} in NWQ area, which leads to greater stability in the throughput. On other hand, One-step-ahead provides high variance through its path, with its median being close the the proposed \ac{rl} approach, in 70 Mbps, and its first and third quartile been between 47 Mbps and 78 Mbps. The Random walk approach shows a small variance on its quartile, although it also shows many outliers. As the behaviour is random, the outliers showed a significant variation of the throughput. However, on average, it manages to maintain a throughput near its median of 46 Mbps.

Figure \ref{fig:gcqheight} illustrates how the different strategies adjusted the UAV heights when flying in the GCQ area. 
As in the NWQ area, we inspect the individual height adaptation to understand how each of the approaches behave in a real path and have an idea of how many adaptation were needed to achieve their respective throughput for the GCQ area.
Here, we observe that our proposed solution maintains a low height when flying near a macro cell deployment maintaining its height at 20 m or 30 m at all times. Also, we note that the Optimal height at each step requires significant changes in the UAV height from step to step for example. The One-step-ahead approach follows the Optimal height and moves up and down 50 times for these 100 steps.
In this example run, the Random walk approach starts at a higher height and keeps moving randomly until move to the lower heights. 
%In this scenario, as the \ac{uav} had some periods where its signal was blocked by buildings, the number of height changes made by the RL solution increased to 34, compared to the NWQ set of measurements. However, it is still a smaller when compared to the non-constant approaches, random, genie, optimal movement. 
\section{Discussion \& Conclusion}
\label{sec:conclusion}
% !TEX root = ../../main.tex
In this paper, we presented a \ac{rl}-based approach to optimise the height at which a mobile cellular-connected \ac{uav} should fly. Our primary objective was to increase the \ac{uav}'s average \ac{qos}. We  evaluated the proposed approach in a generated environment and varied BS density and building density. We also evaluated our approach using a experimental dataset based on real data obtained from a \ac{uav} carrying a smartphone in two locations of Dublin city centre~\cite{measurament}. 
The performance achieved in both scenarios was comparable, where the proposed \ac{rl} approach 
%First thing to notice is that the macro cell provided a better throughput when compared to the small cell deployment for most of the approaches, although the small cell could provide the throughput through all the heights with lower difference between the throughput.
was shown to be successful in both environments, providing an improvement of 6\% compared to other approaches, including the ones that had access to additional priori information about the environment. We  conclude that for low BS density the UAV usually maintain higher \ac{qos} then when compared with higher BS densities. We concluded that the variance of building density when the UAV is flying over them does not change UAV's \ac{qos}.

However, we believe that there is a threshold to be considered when using the proposed solution. For example, if the \ac{uav} need to inform its exact location prior to the flight,  a good approach would be the one proposed by Zhang \cite{trajec-mobile}, where it maintains the lowest possible height through the flight.  However, if the \ac{qos} of the connection is mission-critical \ac{uav} priority and the \ac{uav} can adapt its location during the flight, the \ac{uav} could use the proposed RL solution.
% 

%Extending this evaluation to different datasetscities would be an interesting topic for our future work. 
As a topic for future work, we are interested in evaluating how much energy is associated with the height changes and how to incorporate this factor into the height adaptation decision. An additional challenge that we plan to investigate is how to jointly adapt the horizontal and vertical trajectory of a cellular-connected \ac{uav} in order to improve its \ac{qos}. 
%Another aspect that can be investigated is the choice of which \ac{bs} the \ac{uav} connects to so that the network is not negatively impacted by handover effects. The selection of the connected \ac{bs} can also be optimised in order to increase \ac{qos} in the long term, considering the penalties introduced by frequent \ac{bs} handovers.

%The code used in this paper is public available for the research community in GitHub.
\section*{Acknowledgements}

The research leading to this work is funded, in part, by Science Foundation Ireland (SFI) and the National Natural Science Foundation of China (NSFC) under the SFI-NSFC Partnership Programme Grant Number 17/NSFC/5224 and SFI grant 13/RC/2077 P2.  It was also supported by the Commonwealth Cyber Initiative (CCI).

\bibliographystyle{IEEEtran}
\bibliography{IEEEabrv,main}

% Generated by IEEEtran.bst, version: 1.14 (2015/08/26)
\begin{thebibliography}{10}
\providecommand{\url}[1]{#1}
\csname url@samestyle\endcsname
\providecommand{\newblock}{\relax}
\providecommand{\bibinfo}[2]{#2}
\providecommand{\BIBentrySTDinterwordspacing}{\spaceskip=0pt\relax}
\providecommand{\BIBentryALTinterwordstretchfactor}{4}
\providecommand{\BIBentryALTinterwordspacing}{\spaceskip=\fontdimen2\font plus
\BIBentryALTinterwordstretchfactor\fontdimen3\font minus
  \fontdimen4\font\relax}
\providecommand{\BIBforeignlanguage}[2]{{%
\expandafter\ifx\csname l@#1\endcsname\relax
\typeout{** WARNING: IEEEtran.bst: No hyphenation pattern has been}%
\typeout{** loaded for the language `#1'. Using the pattern for}%
\typeout{** the default language instead.}%
\else
\language=\csname l@#1\endcsname
\fi
#2}}
\providecommand{\BIBdecl}{\relax}
\BIBdecl
\renewcommand{\BIBentryALTinterwordstretchfactor}{4}

\bibitem{release14}
``{3rd Generation Partnership Project Technical Specification Group Radio
  Access Network},'' {3GPP}, Tech. Rep., March 2017.

\bibitem{rami1}
M.~Mozaffari, W.~Saad, M.~Bennis, Y.-H. Nam, and M.~Debbah, ``A tutorial on
  uavs for wireless networks: Applications, challenges, and open problems,''
  \emph{{IEEE Communications Surveys \& Tutorials}}, 2019.

\bibitem{rami2}
M.~Mozaffari, W.~Saad, M.~Bennis, and M.~Debbah, ``Unmanned aerial vehicle with
  underlaid device-to-device communications: Performance and tradeoffs,''
  \emph{{IEEE Transactions on Wireless Communications}}, 2016.

\bibitem{reshape}
M.~M. Azari, F.~Rosas, and S.~Pollin, ``Reshaping cellular networks for the
  sky: Major factors and feasibility,'' in \emph{2018 IEEE International
  Conference on Communications (ICC)}.\hskip 1em plus 0.5em minus 0.4em\relax
  IEEE, 2018, pp. 1--7.

\bibitem{network-adapt}
M.~M. Azari, F.~Rosas, A.~Chiumento, and S.~Pollin, ``Coexistence of
  terrestrial and aerial users in cellular networks,'' in \emph{2017 IEEE
  Globecom Workshops (GC Wkshps)}.\hskip 1em plus 0.5em minus 0.4em\relax IEEE,
  2017, pp. 1--6.

\bibitem{rami3}
R.~Amer, W.~Saad, and N.~Marchetti, ``Mobility in the sky: Performance and
  mobility analysis for cellular-connected uavs,'' \emph{IEEE Transactions on
  Communications}, vol.~68, no.~5, pp. 3229--3246, 2020.

\bibitem{trajectory1}
A.~{Richards} and J.~P. {How}, ``Aircraft trajectory planning with collision
  avoidance using mixed integer linear programming,'' in \emph{Proceedings of
  the 2002 American Control Conference (IEEE Cat. No.CH37301)}, vol.~3, 2002,
  pp. 1936--1941 vol.3.

\bibitem{trajec-mobile}
S.~{Zhang}, Y.~{Zeng}, and R.~{Zhang}, ``Cellular-enabled uav communication: A
  connectivity-constrained trajectory optimization perspective,'' \emph{IEEE
  Transactions on Communications}, vol.~67, no.~3, pp. 2580--2604, 2019.

\bibitem{rival}
U.~Challita, W.~Saad, and C.~Bettstetter, ``Interference management for
  cellular-connected uavs: A deep reinforcement learning approach,'' \emph{IEEE
  Transactions on Wireless Communications}, vol.~18, no.~4, pp. 2125--2140,
  2019.

\bibitem{3d-2}
R.~I. {Bor-Yaliniz}, A.~{El-Keyi}, and H.~{Yanikomeroglu}, ``Efficient 3-d
  placement of an aerial base station in next generation cellular networks,''
  in \emph{2016 IEEE International Conference on Communications (ICC)}, 2016,
  pp. 1--5.

\bibitem{3d-1}
E.~{Kalantari}, H.~{Yanikomeroglu}, and A.~{Yongacoglu}, ``On the number and 3d
  placement of drone base stations in wireless cellular networks,'' in
  \emph{2016 IEEE 84th Vehicular Technology Conference (VTC-Fall)}, 2016, pp.
  1--6.

\bibitem{3d-3}
X.~{Liu}, Y.~{Liu}, and Y.~{Chen}, ``Reinforcement learning in multiple-uav
  networks: Deployment and movement design,'' \emph{IEEE Transactions on
  Vehicular Technology}, vol.~68, no.~8, pp. 8036--8049, 2019.

\bibitem{44}
A.~Al-Hourani, S.~Kandeepan, and S.~Lardner, ``Optimal lap altitude for maximum
  coverage,'' \emph{IEEE Wireless Communications Letters}, vol.~3, no.~6, pp.
  569--572, 2014.

\bibitem{61}
V.~V.~C. Ravi and H.~S. Dhillon, ``Downlink coverage probability in a finite
  network of unmanned aerial vehicle (uav) base stations,'' in \emph{2016 IEEE
  17th International Workshop on Signal Processing Advances in Wireless
  Communications (SPAWC)}.\hskip 1em plus 0.5em minus 0.4em\relax IEEE, 2016,
  pp. 1--5.

\bibitem{62}
V.~V. Chetlur and H.~S. Dhillon, ``Downlink coverage analysis for a finite 3-d
  wireless network of unmanned aerial vehicles,'' \emph{IEEE Transactions on
  Communications}, vol.~65, no.~10, pp. 4543--4558, 2017.

\bibitem{ITUR_2012}
``{Recommendation P.1410-5 "Propagation Data and Prediction Methods Required
  for the Design of Terrestrial Broadband Radio Access Systems Operating in a
  Frequency Range From 3 to 60 GHz"},'' ITU-R, Tech. Rep., 2012.

\bibitem{galkincoverage}
B.~Galkin, J.~Kibilda, and L.~A. DaSilva, ``Coverage analysis for low-altitude
  uav networks in urban environments,'' in \emph{GLOBECOM 2017-2017 IEEE Global
  Communications Conference}.\hskip 1em plus 0.5em minus 0.4em\relax IEEE,
  2017, pp. 1--6.

\bibitem{boris_dic_ant}
------, ``Backhaul for low-altitude uavs in urban environments,'' in \emph{2018
  IEEE International Conference on Communications (ICC)}.\hskip 1em plus 0.5em
  minus 0.4em\relax IEEE, 2018, pp. 1--6.

\bibitem{sutton2018reinforcement}
{Sutton, Richard S and Barto, Andrew G}, \emph{{Reinforcement learning: An
  introduction}}.\hskip 1em plus 0.5em minus 0.4em\relax {MIT press}, {2018}.

\bibitem{reqiba}
{Galkin, Boris and Fonseca, Erika and Amer, Ramy and A. DaSilva, Luiz and
  Dusparic, Ivana}, ``{REQIBA: Regression and Deep Q-Learning for Intelligent
  UAV Cellular User to Base Station Association},'' \emph{{IEEE Transactions on
  Vehicular Technology}}, vol.~71, no.~1, pp. 5--20, {2022}.

\bibitem{nallanathan}
J.~Cui, Z.~Ding, Y.~Deng, A.~Nallanathan, and L.~Hanzo, ``Adaptive
  uav-trajectory optimisation under quality of service constraints: A
  model-free solution,'' \emph{IEEE Access}, vol.~8, pp. 112\,253--112\,265,
  2020.

\bibitem{measurament}
B.~Galkin, E.~Fonseca, G.~Lee, C.~Duff, and M.~Kelly, ``{Experimental
  Evaluation of a UAV User QoS from a Two-Tier 3.6GHz Spectrum Network},'' in
  \emph{IEEE ICC Workshops}, 2021.

\end{thebibliography}

% Comment out if necessary
%\section*{Biographies}
%\input{biography}

\end{document}